\documentclass[reprint,amsmath,amssymb,prl,twocolumn]{revtex4}
\usepackage{amsfonts}    
\usepackage{amssymb}
\usepackage{latexsym}
\usepackage{graphicx}
\usepackage{rotating}
\usepackage{multirow}
\usepackage[usenames,dvipsnames]{color}
\usepackage[active]{srcltx}

 \usepackage{float}
\usepackage[caption = false]{subfig}

\usepackage{hyperref}
\usepackage{color}

\newcommand{\ta}{\tau}

\newcommand{\de}{\delta}

\newcommand{\rar}{\rightarrow}
\newcommand{\lrar}{\leftrightarrow}

\newcommand{\non}{\nonumber}

\begin{document}

\title{Superintegrability of $(2n+1)$-body  choreographies, $n=1,2,3,\ldots, \infty$
 on the algebraic Lemniscate by Bernoulli (inverse problem of classical mechanics)}
\date{\today}

\author{Alexander~V.~Turbiner}
\email{turbiner@nucleares.unam.mx}
\author{Juan Carlos Lopez Vieyra}

\email{vieyra@nucleares.unam.mx}
\affiliation{Instituto de Ciencias Nucleares, Universidad Nacional
Aut\'onoma de M\'exico, Apartado Postal 70-543, 04510 M\'exico,
D.F., Mexico{}}

\begin{abstract}
For one 3-body and two 5-body planar choreographies on the same algebraic Lemniscate
by Bernoulli we found explicitly a maximal possible set of (particular) Liouville integrals, 7 and 15, respectively, (including the total angular momentum),
which Poisson commute with the corresponding Hamiltonian along the trajectory. Thus, these choreographies are particularly maximally superintegrable. It is conjectured that the total number of (particular) Liouville integrals is maximal possible for any odd number of bodies $(2n+1)$ moving choreographically (without collisions) along given algebraic Lemniscate, thus, the corresponding trajectory is particularly, maximally superintegrable.
Some of these Liouville integrals are presented explicitly.
The limit $n \rar \infty$ is studied: it is predicted that one-dimensional liquid with nearest-neighbor interactions occurs, it moves along algebraic Lemniscate and it is characterized by infinitely-many constants of motion.
\end{abstract}

\maketitle
\section{Introduction}

Recently, it was shown~\cite{Turbiner:2020} that the Remarkable Figure Eight trajectory
for the three body Newtonian problem with equal masses, discovered by Moore \cite{Moore:1993} numerically and re-discovered by Chenciner-Montgomery \cite{CM:2000} mathematically, is characterized by 2 global and 5 particular Liouville integrals (constants) of motion. It was found 6 independent functions on phase space, which remain constant in evolution and have a vanishing Poisson bracket with the Hamiltonian along the trajectory. One integral among them is global, this is the total angular momentum while the remaining five are particular integrals, they were found approximately. Each of these particular integrals  was represented as a polynomial of finite degree in the particular Liouville integral of the system of three equal mass bodies moving choreographically on the algebraic lemniscate by Jacob Bernoulli - the so-called FFO model \cite{Fujiwara:2003},
see below - (see \cite{Turbiner:2020} and references therein).

Thus, the Remarkable Figure Eight Newtonian three body trajectory by Moore and
the three body choreography on the algebraic lemniscate by Jacob Bernoulli due to Fujiwara-Fukuda-Ozaki (FFO) \cite{Fujiwara:2003} are two examples of particular maximally superintegrable 3-body systems. This discovery supports the conjecture made in \cite{Turbiner:2013} which asserts that in classical mechanics the existence of a number of (special) closed trajectories is related to the existence of additional, particular constants of motion for these trajectories only (we will call it the {\it T-conjecture}). This conjecture poses the question about a generalization of the  Nekhoroshev theorem, which states that in the case of maximal superintegrability, where all (dimension of phase space minus one) integrals are global, implies that {\it all} bounded trajectories in coordinate space are closed (and periodic) \cite{Nekhoroshev:1972};
for concrete examples, where this theorem holds, see for instance \cite{TTW:2010}.
Its possible generalization can be formulated as follows:
the existence of special (closed) trajectories can be related to the appearance of a certain number of additional, particular integrals (the so-called $\pi$-integrals, see \cite{Turbiner:2013}). In particular, for a maximal possible number of {\it particular} integrals (constants of motion) associated with a certain trajectory the closed periodic trajectory can occur, such a trajectory is called the {\it superintegrable trajectory}.

In the present paper we give arguments to show that (a) choreographies of five, seven and presumably any odd number of bodies on the algebraic lemniscate by Bernoulli are solutions of the systems of coupled Newton equations of motion for a well defined Hamiltonian, and (b) that such choreographies, being a set of closed periodic trajectories of the same form, are maximally particularly superintegrable. For the case of five bodies we find explicitly   15 Liouville particular integrals (the maximal number) and 12 (out of 23) Liouville integrals for the 7-body case.

\section{\it (A) 3-body choreography on the algebraic lemniscate}

\begin{figure}[t]
\includegraphics[trim=0.0in 1.0in 0.0in 1.25in, clip,height=2in]{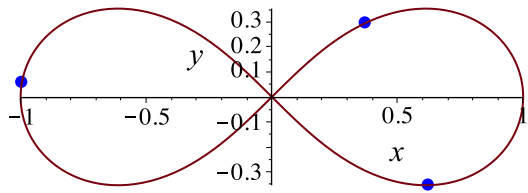}
\caption{\label{fig1} Three equal masses moving on the algebraic Lemniscate by Bernoulli (\ref{lemn}) for $c=1$.}
\end{figure}

As the first step let us review briefly the 3-body choreography on the algebraic Lemniscate proposed in 2003 by Fujiwara et al, \cite{Fujiwara:2003} (see Fig.\ref{fig1}).

\begin{figure}[tbh]
\includegraphics[height=2.25in]{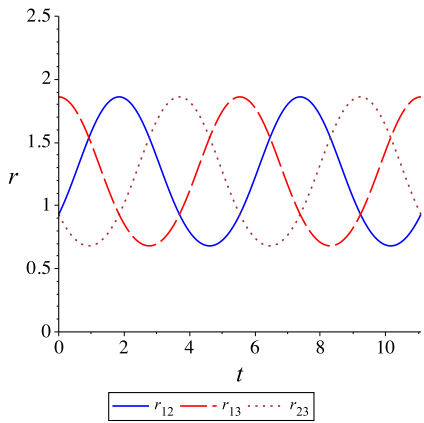}
\caption{\label{fig2}  Time evolution over one period of the relative distances
$r_{12}$ (solid), $r_{13}$ (dashed), $r_{23}$ (dotted) for three equal masses
moving on the algebraic Lemniscate by Bernoulli (\ref{lemn}) for $c=1$.
All three distances oscillate in the same domain  $r_{ij} \in [r^{\rm min},  r^{\rm max}] =
[ \sqrt{2\sqrt{3} -3 }, \sqrt[4]{12}]\, $.}
\end{figure}

\begin{figure}[tbh]
\includegraphics[height=2.25in]{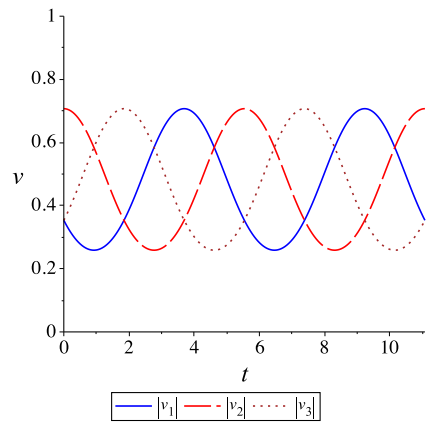}
\caption{\label{fig3}  Time evolution  over one period of the velocities $|{\mathbf v}_1|$ (solid),  $|{\mathbf v}_2|$ (dashed), $|{\mathbf v}_3|$ (dotted) for three equal mass bodies moving on the algebraic Lemniscate by Bernoulli (\ref{lemn}) for $c=1$.  All three velocities oscillate in the same domain
$|{\mathbf v}_i |\in [v^{\rm min},  v^{\rm max}] =
\frac{1}{\sqrt{2}}\ [\frac{\sqrt{3}-1}{2}\, ,\, 1]$\,.}
\end{figure}

The algebraic lemniscate by Jacob Bernoulli (1694) is an algebraic curve of the degree 4 on the $(x,y)$-plane defined by the equation
\begin{equation}
\label{lemn}
    (x^2+y^2)^2\ =\ c^2 (x^2-y^2)\ ,
\end{equation}
where parameter $c$ ``measures" the size of the curve. Without loss of generality
one can put $c=1$. This curve is the common trajectory of three equal masses say $m=1$
chasing each other with the same time delay $\de \tau=\tau/3$, where $\tau$ is the period, see below eq.(\ref{period-3}). It is shown that this trajectory appears as the exact periodic solution of {\it six} coupled Newton equations
\[
  \frac{d^2}{d t^2} {\mathbf x_i} (t)\ =\  -{\nabla_{\mathbf x_i} {\cal V}}\ ,
  \quad \scriptstyle i=1\ldots 3 \ ,
\]
(or {\it four} coupled Newton equations for relative motion after separation of centre-of-mass motion) for 3 point-like bodies subject to pairwise potentials,
\begin{equation}
\label{potential}
 {\cal V}\ =\ \sum_{i<j}^3 \left\{  \frac{1}{4}{\ln r_{ij}^2}\ -\ \beta r_{ij}^2\right\}
 \ \equiv \ \frac{1}{4}\, \ln I_1\ -\ \beta\, I_2\ ,
\end{equation}
with $\beta=\frac{\sqrt{3}}{24}$ at {\it zero} total angular momentum. Here
$r_{ij}=\sqrt{|{\mathbf x}_i -{\mathbf x}_j |^2}\ , j>i=1,2,3$
is the relative distance between bodies $i$ and $j$, where ${\mathbf x}_i, i=1,2,3$ are position vectors. The time evolution of relative distances is shown in Fig.\ref{fig2}: all three relative distances evolve periodically with half-period $\tau/2$. Time evolution  
of the absolute value of the velocity $v_i=|{\mathbf v}_i|\,, i=1,2,3$ is shown in Fig.\ref{fig3}, it is characterized by half-period $\tau/2$ as well.
The first attractive term in (\ref{potential}) is a superposition of three 2-body ${\mathbb R}^2$ Newton
gravitational potentials with the gravitational constant \hbox{$G=1/2$}, while
the second repulsive term is nothing but the moment of inertia (or the square of
the hyperradius in the space of relative coordinates); it represents a pairwise repulsive
harmonic oscillator interaction. This repulsive term dominates at large relative distances,
but the motion occurs at small relative distances, in the attractive region of the potential.
 Following Fujiwara {\it et al.} \cite{Fujiwara:2003} the algebraic lemniscate
(\ref{lemn}) can be  parametrized as
\begin{equation}
\label{param}
 x(t)\ =\ c\, \frac{{\rm sn}(t,k)}{1+{\rm cn}^2(t,k)},\quad
 y(t)\ =\ c\, \frac{{\rm sn}(t,k)\, {\rm cn}(t,k)}{1+{\rm cn}^2(t,k)}\ ,
\end{equation}
see \cite{Fujiwara:2003}, where ${\rm sn}(t,k),\ {\rm cn}(t,k)$ are
Jacobi elliptic functions, $k \in [0,1]$ is the elliptic modulus.
In this parametrization the following relation  ($c=1$) holds:
\begin{equation}
\label{xvrelation}
{\mathbf v}^2(t) + (k^2 - 1/2) {\mathbf x}^2(t) = 1/2 \, ,
\end{equation}
where ${\mathbf x}(t)= (x(t),y(t))$ and ${\mathbf v}(t)= (\dot x(t),\dot y(t))$ are the $2D$ position and velocity vectors.

\begin{figure}[tbh]
\includegraphics[height=2.25in]{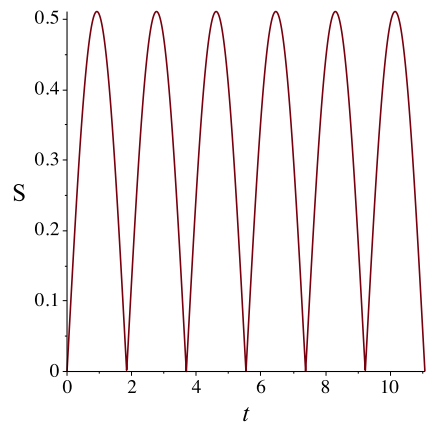}
\caption{\label{fig4}  Time evolution over one period of the area of the triangle $S$
formed by the three equal masses moving on the algebraic Lemniscate by Bernoulli (\ref{lemn}) for $c=1$.  It oscillates in the domain  $S \in [0, S^{\rm max}]$ ,
where $S^{\rm max}=(3/4) r_{ij}^{\rm  min}   = (3/4)\sqrt{2\sqrt{3} -3 }$, corresponding to linear configuration (the Euler line, $S=0$) and to the isosceles configuration ($S^{\rm max}$), respectively.}
\end{figure}

The real period $\tau$, for which $x(t)=x(t+{\tau}), y(t)=y(t+{\tau})$, is given by
\begin{equation}
\label{period}
{\tau} = 4 K(k) = 4 \int_0^1 \frac{dx}{\sqrt{(1-x^2)(1-k^2x^2)}}\ ,
\end{equation}
where $K(k)$ is complete elliptic integral and the parameter $t$ plays the role of the physical time. Then, the evolution of the system is defined by the {\it time}-dependent position vectors
\begin{align}
\label{evolution}
   {\mathbf x}_1(t) &\ =\ ( x(t - {\tau}/{3})\ ,\ y(t - {\tau}/{3})) \ , \non
\\[3pt]
   {\mathbf x}_2(t) &\ =\ (x(t)\ ,\ y(t) ) \ ,
\\[3pt]
   {\mathbf x}_3(t) &\ =\ ({x}(t + {\tau}/{3})\ ,\ y(t + {\tau}/{3}) )\ ,\non
\end{align}
for the first, second and third bodies, respectively. By straightforward calculation
using Maple 18 in 20-digit arithmetic one can check that the center-of-mass is conserved
\[
  {\mathbf X}_{\rm CM}(t) = {\mathbf x}_1 + {\mathbf x}_2 + {\mathbf x}_3 =0\ ,
\]
and fixed
\[
  {\mathbf V}_{\rm CM}(t) = {\mathbf v}_1 + {\mathbf v}_2 + {\mathbf v}_3 =0\ ,
\]
if the elliptic modulus takes the value
\[
 k_0^2= \frac{2+\sqrt{3}}{4} = \left(\frac{1+\sqrt{3}}{2\sqrt{2}}\right)^2 \
\]
only, see \cite{Fujiwara:2003}, it corresponds to the period of the motion
\begin{equation}
\label{period-3}
 \tau=11.07225258147507023547\ .
\end{equation}
The value of the elliptic modulus is the root
of the equation \cite{Fujiwara:2004}
\begin{equation}
\label{3body-k}
        \mathrm{dn}\left(\frac{5K(k_0)}{3}\right)\ =\ \frac{1}{\sqrt{2}}\ .
\end{equation}
During the period the system passes subsequently the linear configurations (the Euler line), where the area of the triangle formed by the three bodies is zero,
to the isosceles configurations, where the area of the triangle becomes maximal, at times $t=p\tau/12\,, p=0,1,\ldots 12$ (see Fig.\ref{fig4}).

It can be checked that the following functions (polynomial in coordinates and velocities)  become constants of motion under the evolution (\ref{param}), (\ref{evolution}), {\it i.e.} along the trajectory (\ref{lemn}), see \cite{Fujiwara:2003,Turbiner:2020}:
\begin{align}
 L& =\sum {\mathbf x}_i \times {\mathbf v}_i = 0\ ,  \non \\ 
 E & ={T} + {\cal V}=\frac{1}{4}\log\left(\frac{3\sqrt{3}}{2}\right)\ ,  \non \\ 
 I_1 &= r_{12}^2 \,r_{13}^2 \,r_{23}^2 =   \frac{3\sqrt{3}}{2}\ ,  \label{integrals3b} \\
 I_2 &= 3\sum {\mathbf x}_i^2 = \sum_{i<j} r_{ij}^2 = 3\sqrt{3}\ , \non \\  
 {\tilde {\cal T}} &= {\mathbf v}_1^2 \, {\mathbf v}_2^2 \,{\mathbf v}_3^2  =   \frac{1}{128}\ , \non \\
 J_1 &= {\mathbf v}_1^2 +  \frac{1}{9} (k_0^2 - \frac{1}{2})
 \left( 2\, r_{12}^2 + 2\, r_{13}^2  -\, r_{23}^2 \right) = \frac{1}{2}\ , \non \\
 J_2 &= {\mathbf v}_2^2 +  \frac{1}{9} (k_0^2 - \frac{1}{2} )
 \left( 2\, r_{12}^2  -\, r_{13}^2  + 2\, r_{23}^2 \right) = \frac{1}{2}\ , \non \\
 J_3 &= {\mathbf v}_3^2 +  \frac{1}{9} (k_0^2 - \frac{1}{2})
 \left( -\, r_{12}^2 + 2\, r_{13}^2  +2\, r_{23}^2 \right) = \frac{1}{2} \ ,  \non \\
 \sum^3 J_i &= 2\ {T} + \frac{1}{3}(k_0^2 -  \frac{1}{2})I_2\ , \non
\end{align}
\label{int-3}

\noindent
where  $L$ and $E$ are the total angular momentum and total energy respectively, which
are global integrals of motion; $I_2$ is the moment of inertia, see above. The fact that $I_2$ is a constant of motion does not imply that the motion is a relative equilibrium (contrary to the Saari's conjecture \cite{Saari:1970}
for the ${\mathbb R}^3$ Newtonian Gravity $n$-body problem). It is worth to emphasize
that there exists a surprising duality between $\tilde{\cal T}$ which
is the product of velocities squared  (see figure (\ref{fig3})) and the product of relative distances squared
$I_1$. This duality is also observed between the kinetic energy
$T=\frac{1}{2}\sum {\mathbf v}_i^2 $ and the moment of inertia $I_2$.
Likely, it reflects a hidden symmetry of the system. Similar duality occurs for all $(2n+1)$ choreographies.

It is interesting to see that if we use the relative distances squared $(r^2_{12},r^2_{13},r^2_{23})$ as coordinates, the motion of the system takes place on a planar elliptic curve: the intersection of the surfaces defined by the integrals $I_1=3\sqrt{3}/2$ and $I_2=3\sqrt{3}$ (see (\ref{integrals3b})), these integrals play a role of elliptic invariants.

Seven of above functions (5) are functionally (algebraically)
independent. Let us choose  the set $\{L, I_1, I_2, {T}, \tilde{T}, J_1,
J_2\}$\ , while the Hamiltonian
\begin{equation}
\label{HF}
{\cal H}={T} + \frac{1}{4}\, \ln I_1 - \frac{\sqrt{3}}{24}\, I_2\ ,
\end{equation}
see (\ref{potential}), is made from particular integrals $T$ (the kinetic energy),
$I_1$ and $I_2$. It can be shown explicitly that all seven functions have vanishing
Poisson brackets with the Hamiltonian on the algebraic Lemniscate (\ref{lemn}), {\it i.e.} they are {\it particular} Liouville integrals.
Thus, 3-body choreographic motion (\ref{evolution}) along the algebraic lemniscate  (\ref{lemn}) with pairwise potential (\ref{potential}) is maximally particularly superintegrable.

\bigskip

\section{\it (B)  5-body choreography on the Lemniscate}

\begin{figure}
\subfloat[]{\includegraphics[width = 1.7in]{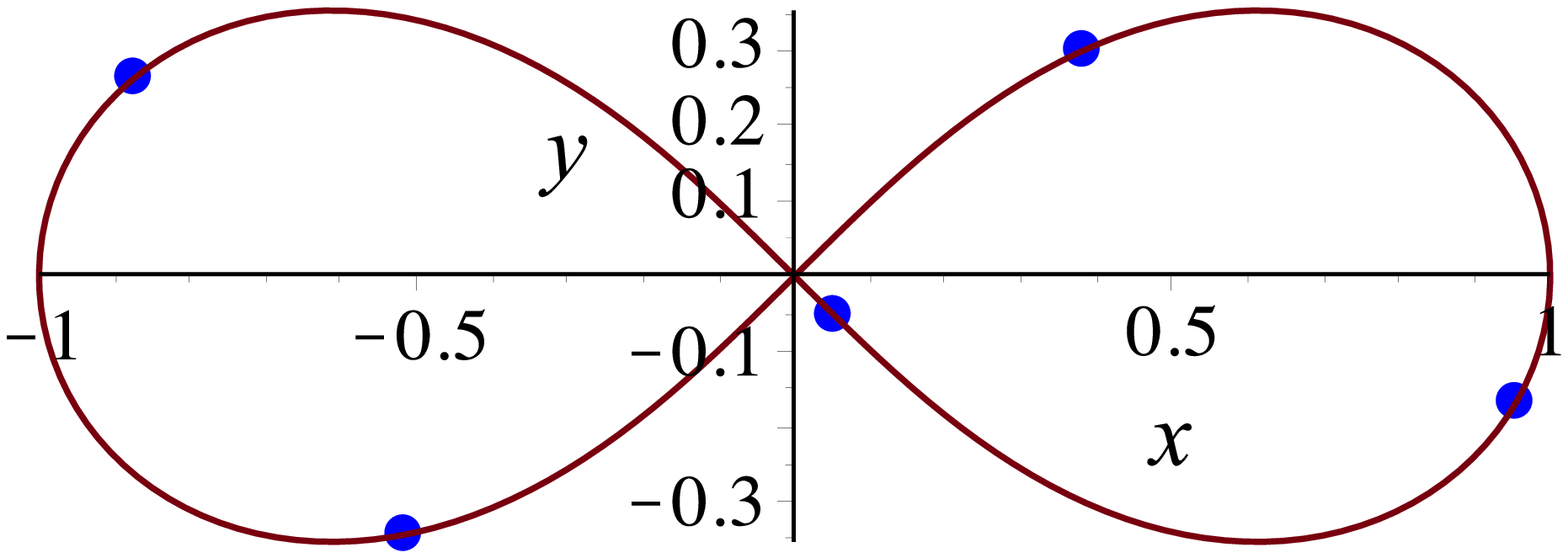}}
\subfloat[]{\includegraphics[width = 1.7in]{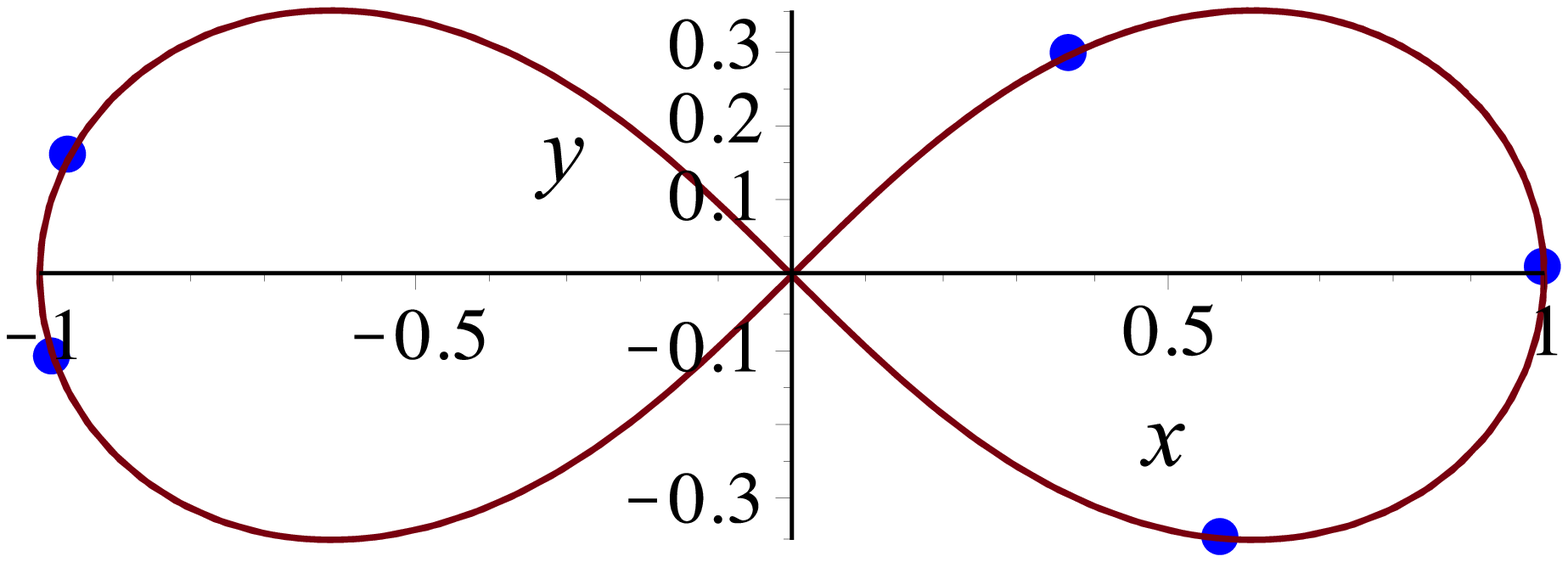}}
\caption{\label{fig5} Five  equal masses  moving on the algebraic Lemniscate by Bernoulli (\ref{lemn}) for $c=1$ (snapshots), (a) for $k_1$, and (b) for $k_2$ (see text).}
\end{figure}

The question about the existence of 5-body choreographies
of equal masses moving on the {\it same} algebraic lemniscate (\ref{lemn})
was explored in Fujiwara {\it et al.} \cite{Fujiwara:2004}.  It was
discovered the existence of two 5-body choreographies which correspond to two different
values of elliptic modulo $k_{1,2}$, thus, two different sets of initial data (see figure (\ref{fig5})).

Recently, it was shown \cite{JC:2019} that the both 5-body choreographies are two solutions of the ten coupled Newton equations
\[
  \frac{d^2}{d t^2} {\mathbf x_i} (t)\ =\  -{\nabla_{\mathbf x_i} {\cal V}}\ ,
  \quad \scriptstyle i=1\ldots 5 \ ,
\]
where ${\mathbf x}_i, i=1,2,3$ are position vectors, (or eight coupled Newton equations for relative motion after separation of centre-of-mass motion) corresponding to pairwise interactions between bodies of the type (\ref{potential}) of the superposition of logarithmic term and quadratic term. For the first solution the first logarithmic term contains {\it nearest neighbors} interactions, while for the second solution it contains the {\it next-to-nearest neighbors}) interactions. The second term in both cases represents the pairwise repulsive harmonic oscillator potentials among all particles.

In \cite{JC:2019} ten independent conserved quantities were found including the energy and total angular momentum. However, for the 5-body choreographies (moving on the algebraic lemniscate) in order to be maximally particularly superintegrable, it should occur $15$
constants of motion (or equivalently, 15 global and/or {\it particular} Liouville integrals). These extra constants/integrals are found and presented below.

To define 5-body choreography on the lemniscate let us place five unit masses
on the curve (\ref{lemn}) with equal time-delay $\ta/5$ between them.
The position of each particle is given by the time-dependent vectors
\begin{align}
  {\mathbf x}_1(t) &\ =\ ( x(t{ -2{\tau}/5})  \,,  y(t-2{\tau}/5) ) \ ,\non \\
  {\mathbf x}_2(t) &\ =\ ( x(t{ -{\tau}/5})   \,,  y(t-{\tau}/5) )  \ ,\non \\
  {\mathbf  x}_3(t) &\ =\ ( x(t)\,, y(t) )     \ ,              \label{xoft5}           \\
  {\mathbf  x}_4(t) &\ =\ ( x(t{ +{\tau}/5})   \,,  y(t+{\tau}/5) ) \ ,\non \\
  {\mathbf  x}_5(t) &\ =\ ( x(t{ +2{\tau}/5})  \,,  y(t+2{\tau}/5) ) \ ,\non
\end{align}
where $x(t)$ and $y(t)$ are given by the Lemniscate's parametrization
(\ref{param}) and ${\tau}=4K(k)$ is the period of the motion. It can be easily
found that such a period is defined by the condition that the center of
mass of the system will remain fixed, {\it i.e.}
 \[
  {\mathbf X}_{ \sc   CM}(t)\ =\
  {\mathbf x}_1(t) + {\mathbf x}_2(t) + {\mathbf x}_3(t) + {\mathbf x}_4(t) + {\mathbf x}_5(t)\ =\ 0\ .
  \]
  \[
  {\mathbf V}_{ \sc   CM}(t)\ =\
  {\mathbf v}_1(t) + {\mathbf v}_2(t) + {\mathbf v}_3(t) + {\mathbf v}_4(t) + {\mathbf v}_5(t)\ =\ 0\ .
  \]
This condition is satisfied only for two values of the elliptic modulus:
\[
 k^2\ =\ \begin{cases}
    0.65366041395477321345\ =\ k_1^2 \ ,\\
    0.99764373603161323509\ =\ k_2^2 \ .\\
\end{cases}
\]
These two values are roots of the equations \cite{Fujiwara:2004}
\begin{equation}
\label{5body-k}
 \mathrm{dn}\left(\frac{7K(k_1)}{5}\ ,\ k_1\right)\ =\ \frac{1}{\sqrt{2}}\ ,\
 \mathrm{dn}\left(\frac{9K(k_2)}{5}\ ,\ k_2\right)\ =\ \frac{1}{\sqrt{2}}\ ,
\end{equation}
respectively. The corresponding periods $\tau(k)=4K(k)$, see (\ref{period}), are
\[
 \tau(k)\ =\ \begin{cases}
  8.04877705220746848433 \ \ \mbox{for }\ k_1 \ , \\
 17.65458226059668736520 \ \ \mbox{for }\ k_2 \ , \\
\end{cases}
\]
cf.(\ref{period-3}).

\begin{figure}
\subfloat[]{\includegraphics[width = 1.7in]{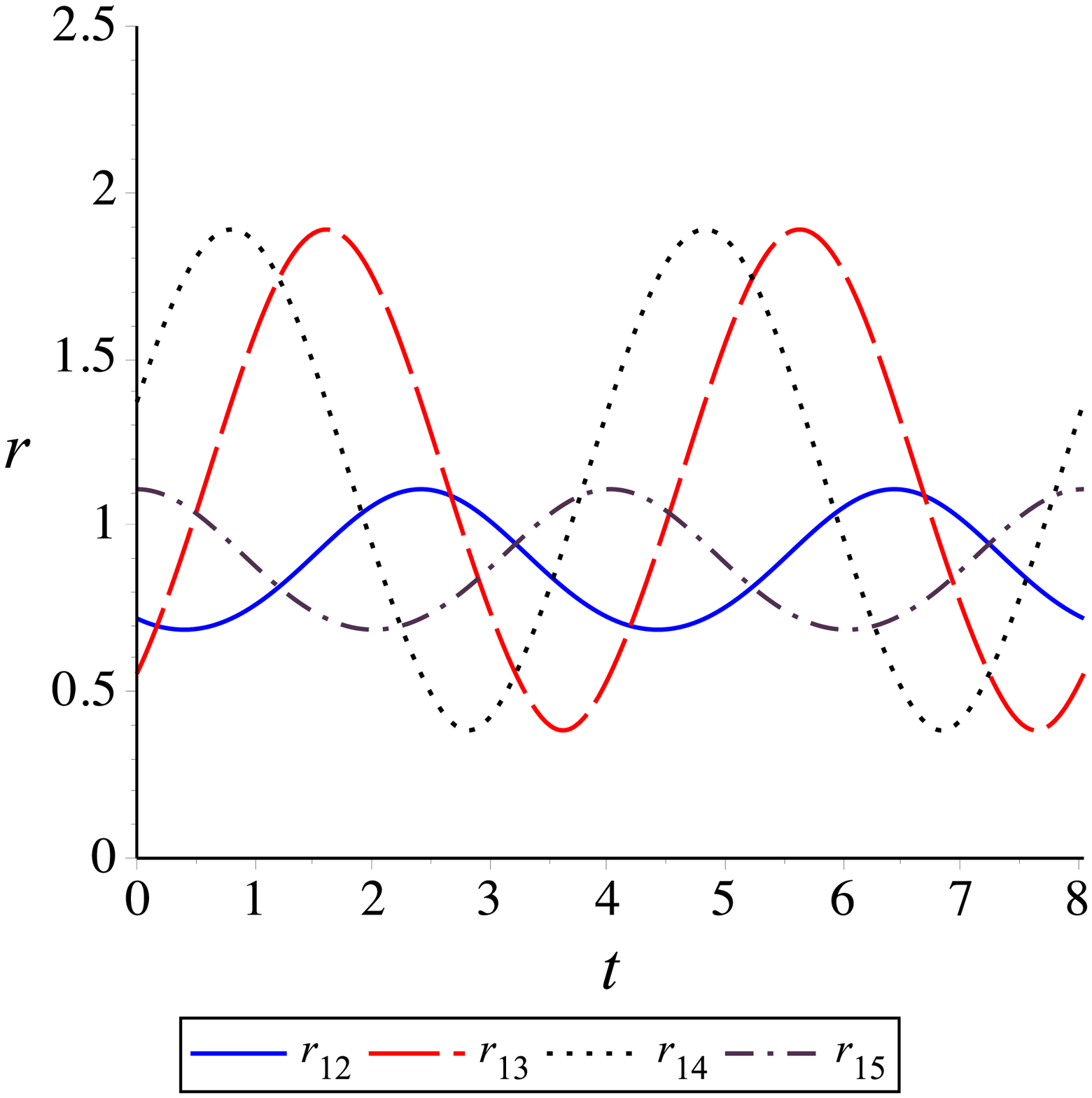}}
\subfloat[]{\includegraphics[width = 1.7in]{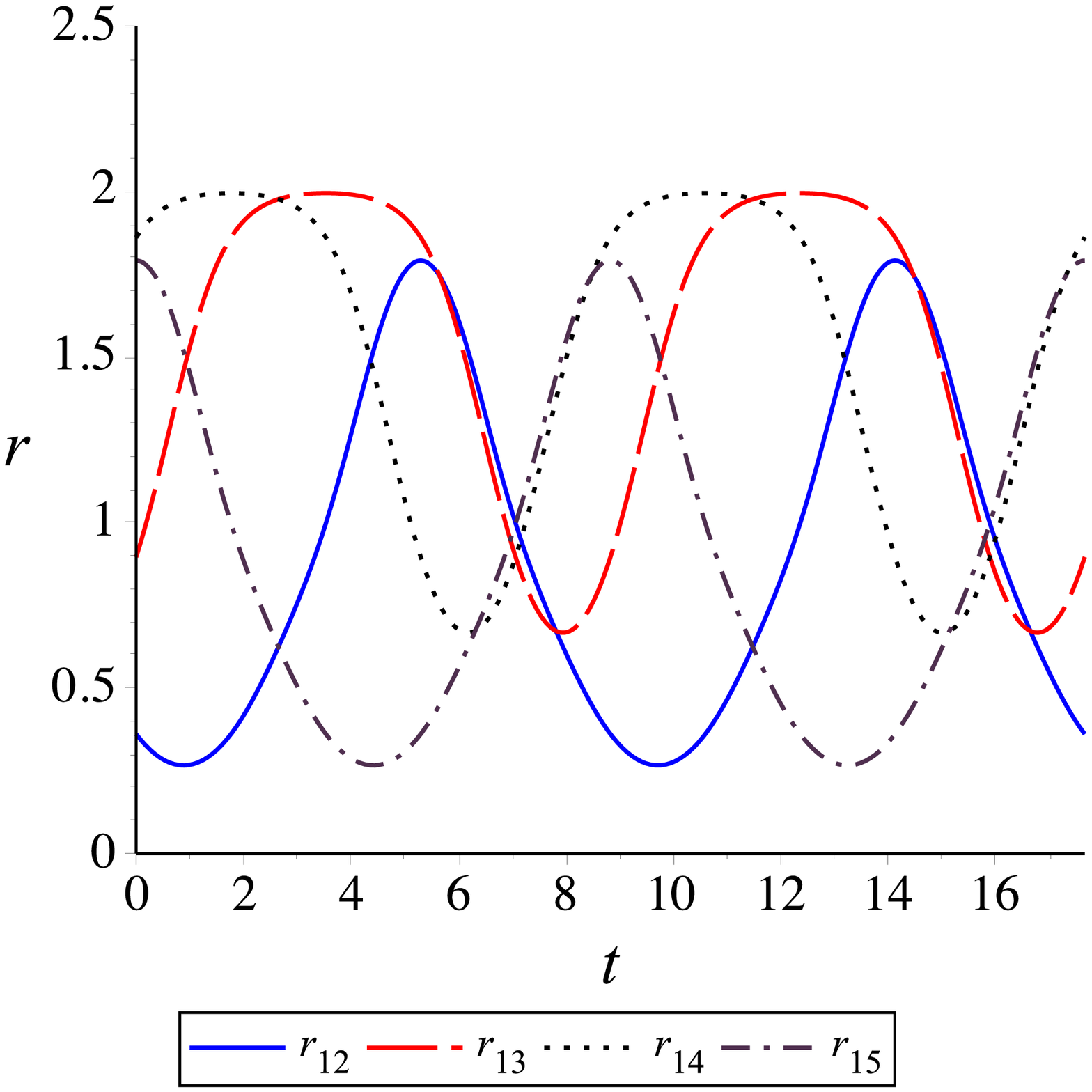}}
\caption{\label{fig6} Time evolution of (some) relative distances $r_{12},r_{13},r_{14},r_{15}$,
 for five  equal masses  moving on the algebraic Lemniscate by Bernoulli (\ref{lemn})
for $c=1$, (a) for $k_1$, and (b) for $k_2$ (see text).}
\end{figure}

\begin{figure}
\subfloat[]{\includegraphics[width = 1.7in]{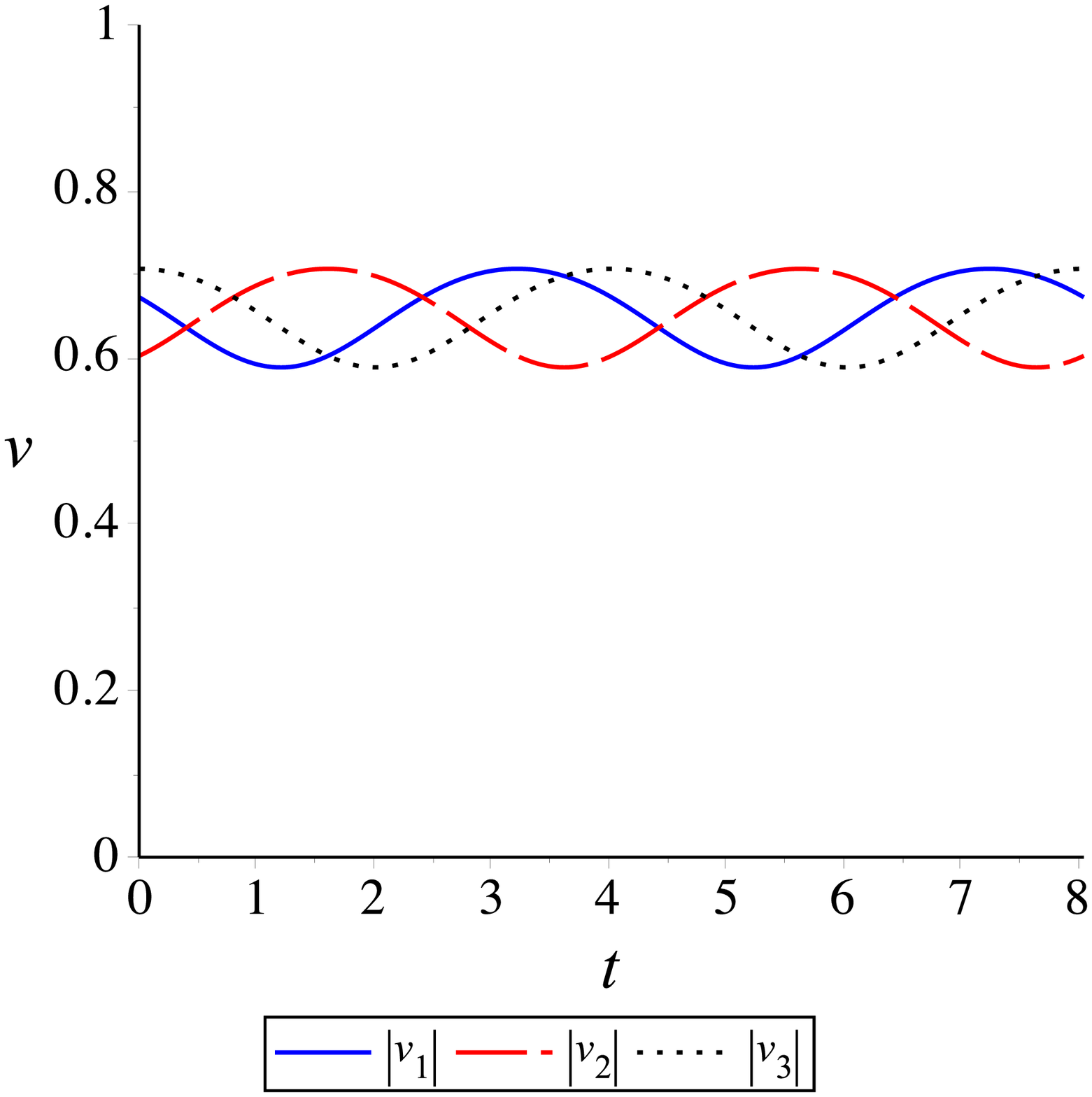}}
\subfloat[]{\includegraphics[width = 1.7in]{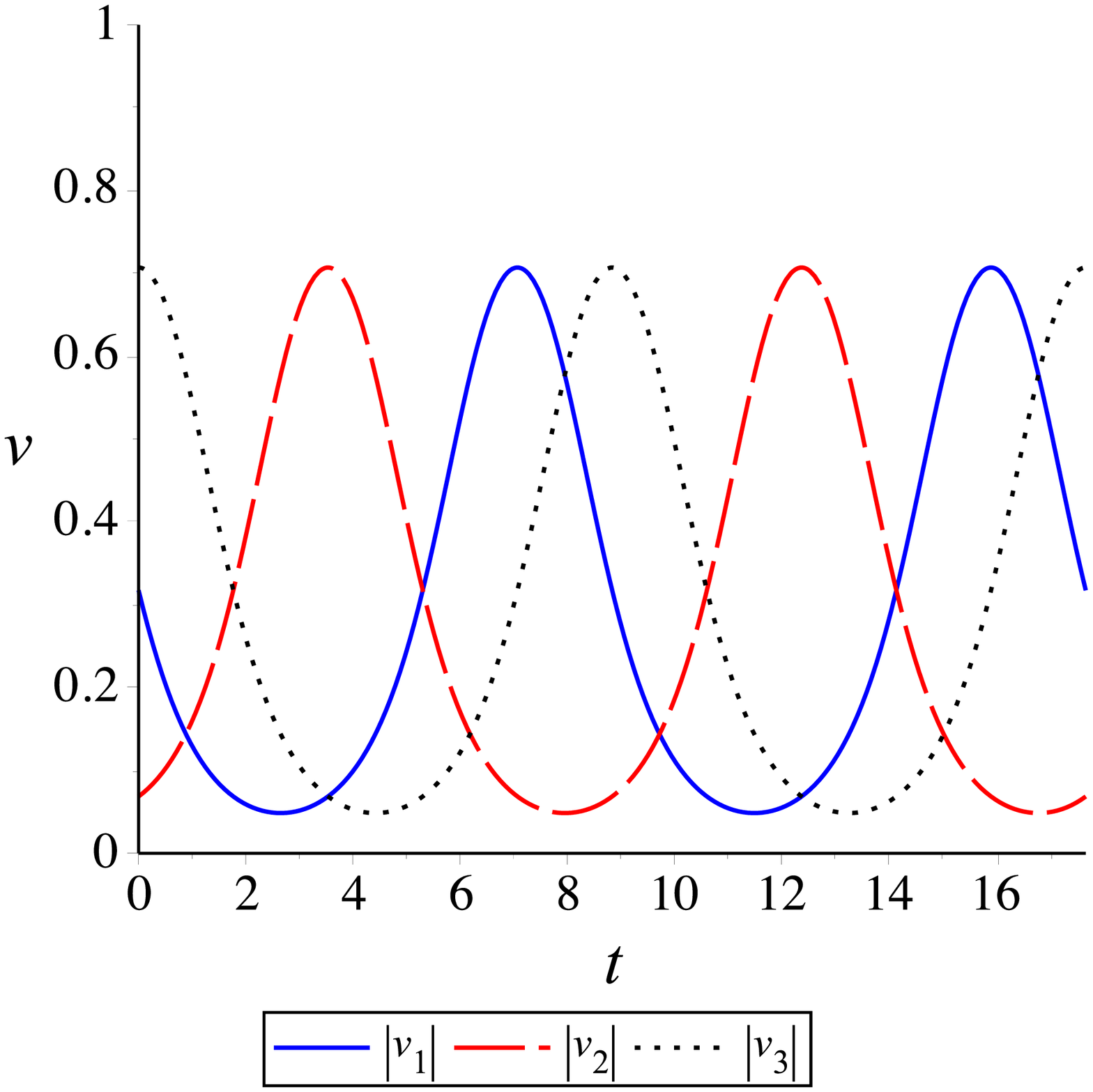}}
\caption{\label{fig7} Time evolution of velocities  $|{\mathbf v}_1|,|{\mathbf v}_2, |{\mathbf v}_3| $, for five equal masses moving on the algebraic Lemniscate by Bernoulli (\ref{lemn}) for $c=1$: (a) for $k_1$, and (b) for $k_2$ (see text). In both cases (a) and (b) the maximal velocity $|{\mathbf v}_{\rm max}| = 1/\sqrt{2}$
when a body passes through the point of self-intersection of the lemniscate at
${\mathbf x}=0$ (see relation (\ref{xvrelation})).}
\end{figure}

The time evolution of the relative distances $r_{ij}$ for both 5-body choreographies at $k_{1,2}$ is shown in Fig.\ref{fig6}. It is periodic and for both cases the period is half of the total period. Comparison of these evolutions shows that for the choreography corresponding to the $k_2 > k_1$ the relative distances are in a broader domain, {\it i.e.} two particles approach closer/further than in the choreography with the elliptic module $k_1$. However,  the choreography corresponding to $k_1$ the motion looks more uniform, smoother. This is seen on Fig.\ref{fig7}: the domain of variation of the absolute value of the velocities for $k_2$ is broader, in particular, the absolute value of the velocity of the body $|{\mathbf v}_i|\,, i=1,2\ldots 5$ can reach rather small value at some moments in the evolution (see below). For both $k$'s the absolute value of the velocities are evolved periodically with period equal to half of the total period.

Apart from the total energy $E$ and the total angular momentum $L$ of the
system, which are global integrals, there exist three special, velocity-independent
constants of motion along the trajectory, which are different for $k_{1,2}$:

{\small

\begin{align}
 I_1^{(5)} &=
  \begin{cases}
     \resizebox{0.7\hsize}{!}{$
 r_{12}^2 r_{23}^2 r_{34}^2 r_{45}^2 r_{15}^2 = 0.26362178303408707110$}\ ,  \\ (\mbox{for}\,k_1)\\[7pt]
    \resizebox{0.7\hsize}{!}{$
 r_{13}^2 r_{35}^2 r_{25}^2 r_{24}^2 r_{14}^2 =30.760801541637359790$}\ ,   \\ (\mbox{for}\,k_2)\\
\end{cases} \label{I15b}
\end{align}

\begin{align}
 I_2^{(5)} &\ =
  \begin{cases}
   \label{integrals5bset1}
   \resizebox{0.7\hsize}{!}{$
  r_{12}^2 + r_{23}^2 + r_{34}^2 + r_{45}^2 + r_{15}^2 = 4.0517817845468308414$} \ ,   \\  (\mbox{for}\,k_1)\\[7pt]
  \resizebox{0.7\hsize}{!}{$
 r_{13}^2 + r_{35}^2 + r_{25}^2 + r_{24}^2 + r_{14}^2 = 12.515257719766335417$} \ ,   \\  (\mbox{for}\ k_2)\\
\end{cases}
\end{align}
\begin{align}
I_{\rm HR}^{(5)} &= 5\sum^5 {\mathbf x}_i^2= \sum^5_{i<j} r_{ij}^2 = \begin{cases}
 11.995383205775537457 \ ,   \\ (\mbox{for}\ k_1)\\
 17.975523091392961251\ ,   \\ (\mbox{for}\ k_2)\\
\end{cases}  \label{IHR5b}
\end{align}
}
\hskip -0.15cm
Let us label bodies on the lemniscate as 1,2,3,4,5.
The functions $I_1^{(5)}(k_1), I_2^{(5)}(k_1)$ contain dependences on the
nearest neighbors $(12),(23),(34),(45),(51)$, while $I_1^{(5)}(k_2), I_2^{(5)}(k_2)$ depend on next-to-nearest neighbors $(13),(24),(35),(41),(52)$.
The variable $I_{\rm HR}^{(5)}$ is nothing but the moment of inertia, or the hyperradius squared in the space of relative distances. It is the sum of the squares of all 10 different relative distances among the bodies.
 The  Kinetic Energy is  also a {constant of motion}.
\begin{equation}
\label{Ekin5b}
{T}= \frac{1}{2}\sum^5_{i=1} {\mathbf v}_i^2 = \begin{cases}
 1.0656784451054396 \  \ (\mbox{for}\,k_1) \ , \\
 0.35545935316766729 \ \ (\mbox{for}\,k_2) \ .  \\
\end{cases}
\end{equation}
Assuming the existence of pairwise interactions only - it leads to a natural guess for the form of the potential
\begin{equation}
\label{fivebpotential}
 {\cal V}\ =\ {\cal V}(I_1^{(5)},I_2^{(5)},I_{\rm HR})\ =\ \alpha\ \log
 I_1^{(5)} + a\ I_2^{(5)} -\beta\ I_{\rm HR}^{(5)}\ .
 \end{equation}
The requirement, that ${\cal V}$ satisfies the coupled Newton equations with evolution (\ref{xoft5}),
leads to the condition $a=0$ and
\begin{align}
 \alpha_1 &\ =\ \frac{1}{4}\ ,\ \beta_1 = 0.015366041395477321360 \ ( \mbox{for}\, k_1)\ ,  \label{V15b}\\
 \alpha_2 &\ =\ \frac{1}{4}\ ,\ \beta_2 = 0.049764373603161323382 \ ( \mbox{for}\  k_2)\ . \label{V25b}
\end{align}
This result confirms that such choreographies are true solutions of the Newton equations
of motion for  potentials of the form (\ref{fivebpotential}). It is worth emphasizing that
both 5-body choreographies take place on the same algebraic lemniscate as trajectory but correspond, in fact, to different potentials. However, the evolution occurs at small relative distances where potentials almost coincide!

In addition to the above-mentioned constants of motion ($E, {L}, I_1^{(5)}, I_2^{(5)}, I_{\rm HR}^{(5)},T$), where in particular, the total energy $E$ takes values
\[
 E\ =\ {\cal T} + {\cal V}\ =\\
 \begin{cases}
0.54804692944384581934 \ \ (\mbox{for}\ k_1)\ ,
 \\
0.31747900688996754830 \ \ (\mbox{for}\ k_2)\ ,
 \end{cases}
\]
it can be shown that the following functions are also constants of motion on the algebraic lemniscate (\ref{lemn}):
%
\widetext
{\small
\begin{align}
\tilde{\cal T} &\ =\
{\mathbf v}_1^2{\mathbf v}_2^2{\mathbf v}_3^2{\mathbf v}_4^2{\mathbf v}_5^2\  =\
\begin{cases}
                           0.01349945192046077\\ (\mbox{for}\ k_1)\ ,\\
                           1.121985660295881\times 10^{-7} \\ (\mbox{for}\ k_2)\ ,
\end{cases}
 \end{align}
}
and
\begin{align}
J_i(k_{1,2})  &\ =\
{\mathbf v}_i^2 + \frac{1}{25}\left(k_{1,2}^2 - \frac{1}{2} \right) \non \times \\
&
\left( b^{(i)}_1 r_{12}^2+ b^{(i)}_2 r_{13}^2
          + b^{(i)}_3 r_{14}^2+ b^{(i)}_4 r_{15}^2  + b^{(i)}_5 r_{23}^2
          + b^{(i)}_6 r_{24}^2 + b^{(i)}_7 r_{25}^2+ b^{(i)}_8 r_{34}^2
          + b^{(i)}_9 r_{35}^2+ b^{(i)}_{10} r_{45}^2 \right)\ =\ \frac{1}{2}\ , \\
          &\ i=1\ldots 5\ . \non
\end{align}
\noindent
where the coefficients $b_{j}^{(i)}$ are given in Table~\ref{bsinJsk1} for $k_1$ and in Table~\ref{bsinJsk2} for $k_2$. It can be checked that $\sum_{i=1}^5\, b_m^{(i)}=5$ for all $m=1,2,..10$. Thus, there is a rather obvious constraint relating the $J$'s functions,
\[
 \sum_{i=1}^5 J_i(k_{1,2})\ =\ 2\ {T}\  +\ \frac{1}{5}
        (k_{1,2}^2- \frac{1}{2})   I_{\rm HR}^{(5)}\ ,
\]
hence, four out of the five $J_i$ functions are independent.

\begin{center}
\begin{table}
\caption{\label{bsinJsk1}}
 {\setlength{\tabcolsep}{0.2cm} \renewcommand{\arraystretch}{1.5}
 \resizebox{0.9\textwidth}{!}{%
\begin{tabular}{|l|lllll|} \hline
$k_1$          &  \hspace{30pt} $i=1$      &   \hspace{30pt}   $i=2$  &  \hspace{30pt} $i=3$      &  \hspace{30pt}   $i=4$    & \hspace{30pt} $i=5$     \\
 \hline
$b^{(i)}_1$    &  -120.82665462346670521 &    -105.79843342244230706 &   -301.42493095562944939 &      188.39841597117506356 &    208.53065966272645523            \\
$b^{(i)}_2$    &  119.07683211113850687 &     46.697069310114807042 &    256.31875981507362749 &         -36.816261165116034132 &   -29.607543418404885011         \\
$b^{(i)}_3$    &  -5.4625813233174444168 &    12.320149654885246091 &    42.958776604400309648 &         -35.273623890684305681 &   -50.187580739914778483         \\
$b^{(i)}_4$    &  483.10462073678451462 &     4.7252178982447825712 &    1237.6914672916637170 &         34.144264368682360500 &          162.60584062877625857    \\
$b^{(i)}_5$    &  282.46864823732293661 &     71.127628469487035041 &    587.17449683780351567 &         58.773801159881296225 &          179.09483665963171711    \\
$b^{(i)}_6$    &  -33.649147614786570551 &    15.509128106068623884 &    -36.236217954127655781 &        -30.308010539225994098 &  -48.801212615020519077          \\
$b^{(i)}_7$    &  121.78083703056524211 &     25.303091050123496630 &    422.35784300796196757 &         -31.068501021324015278 &  -37.615388539926770799          \\
$b^{(i)}_8$    &  -353.58174216822280277 &    -18.703315813976036318 &   -1309.5044069026253110 &        43.801044067886570783 &          60.127456583976772490    \\
$b^{(i)}_9$    &  108.69890919673171619 &     48.806494420763905363 &    301.22659369358541570 &         -55.768924443363832956 &  -58.736576114775790088          \\
$b^{(i)}_{10}$ &  -337.05710071113001903 &    -105.87689989929105649 &   -406.09890140822180453 &        68.249775279761516030 &          43.260443090388281504    \\
\hline
\end{tabular}}}
\end{table}
\end{center}

\begin{center}
\begin{table}
\caption{\label{bsinJsk2}} {\setlength{\tabcolsep}{0.2cm} \renewcommand{\arraystretch}{1.5}
 \resizebox{0.9\textwidth}{!}{%
\begin{tabular}{|l|lllll|} \hline
$k_2$          &  \hspace{30pt} $i=1$      &   \hspace{30pt}   $i=2$  &  \hspace{30pt} $i=3$      &  \hspace{30pt}   $i=4$    & \hspace{30pt} $i=5$     \\
 \hline
$b^{(i)}_1$    &  -61.027100307585863444   &   -17.222850897915190680   &   -511.94270481188472463  &    -25.341131013293214474  &   -122.29941827189394911     \\
$b^{(i)}_2$    &  -0.55282022124686601891  &   3.3268107852559286347    &   420.73598551783063935   &    0.44577751815143152866  &   9.3802959329962503441      \\
$b^{(i)}_3$    &  33.985705553498231158    &   10.173778534367062720    &   -576.75384234875359425  &    1.8893163715313649609   &   48.164237569326552191      \\
$b^{(i)}_4$    &  -26.198188939530811062   &   -5.1656454117960040408   &   -103.59729843163666628  &    -4.5848089247250392683  &   3.8590109120477812307      \\
$b^{(i)}_5$    &  3.2385450799318512530    &   2.5551003803126183551    &   -388.11974128209706643  &    0.22602438154317402832  &   -17.713705937633232610     \\
$b^{(i)}_6$    &  -91.671798678128626528   &   -23.512301517684227553   &   543.59326681675861177   &    -14.126568592308682405  &   -130.34279788542874035     \\
$b^{(i)}_7$    &  17.850234812506293354    &   5.4460925836802282634    &   232.65341630350126493   &    3.7070309607929470383   &   -20.145091389575698176     \\
$b^{(i)}_8$    &  7.6902387946337010174    &   5.3979109151920249436    &   -585.67468059952705546  &    4.7027364153384076972   &   -7.8697278149929396425     \\
$b^{(i)}_9$    &  -7.2933155867104667699   &   2.1998000583101618965    &   -62.658943145935205850  &    -4.0506557851221997180  &   73.586651661155806121      \\
$b^{(i)}_{10}$ &  -76.579699381004642560   &   -37.383379446092551132   &   271.59817931038408011   &    -7.1322099769903142511  &   -189.31984994548931759     \\
\hline
\end{tabular}}}
\end{table}
\end{center}
Note that in a similar way as for 3-body choreography for both 5-body choreographies the quantities ${\tilde{\cal T}}$ and $I_1^{(5)}$ play the role of dual quantities $r^2_{i,i+1} \lrar v^2_i$, as well as for $T$ and $I_{2}^{(5)}$. This duality appears as a basic property for all many-body choreographies we have studied. The interesting fact is that the quantity ${\tilde{\cal T}}$, the product of velocities squared, takes a rather small value for the choreography with elliptic modulus close to one, i.e. $k_2$. It indicates that at some moment in evolution one (or two) body has a small velocity (see Fig. \ref{fig7}). This phenomenon occurs at the end points of the lobes of the lemniscate, where two bodies approach to each other.

Remaining five constants of motion are searched among superpositions in relative distances squared,
\begin{align}
  I_2^{(5)} &\ =\ r_{13}^2 + a r_{34}^2 + b r_{45}^2 + c r_{15}^2 \ =\
\begin{cases}
  -4.9893731063757473516 \ \ (\mbox{for}\,k_1)\ ,\\
 \phantom{+}4.5923978945218266705 \ \ (\mbox{for}\,k_2)\ ,\\
\end{cases}
\\
 &
\begin{array}{lllr}
\hskip -.6cm
\mbox{where}\ \ a = -1\ ,\ &\
 b = -6.1175316692890884736 \ ,&\
 c = a &               \ \ (\mbox{for}\,k_1)\ , \\
 a    =  0.89783594223578596069 \ ,&\
 b    =  0.11755596103339186432\ ,&\
 c    =  a & \ \ (\mbox{for}\,k_2)\ ,
\end{array}
 \\[10pt]
 I_3^{(5)} &\ =\  r_{12}^2 + a r_{14}^2 +  r_{34}^2 + b r_{45}^2 + c r_{15}^2\ =\
 \begin{cases}
   3.8685701539367930365 \ (\mbox{for}\,k_1)\ , \\
   5.0629405424126776331 \ (\mbox{for}\,k_2)\ , \\
\end{cases} \\
  &
   \begin{array}{lllr}
  a = 0.19540670475985874650 \ , &  b  = 1+a  \ , &     c  = 1+a   &  (\mbox{for}\,k_1)\ ,  \\
  a = 1.2815912545379188666 \ ,   &  b  = -0.15065869157919538355 \ , &  c  = b   &  (\mbox{for}\,k_2)\ ,  \non
  \end{array}
  \\[10pt]
  I_4^{(5)} &=  r_{12}^2  + a r_{24}^2 +  r_{45}^2 + b r_{15}^2 =
 \begin{cases}
  4.9893731063757975999  \ (\mbox{for}\,k_1)\ , \\
  5.1149633006284987259  \ (\mbox{for}\,k_2)\ , \\
\end{cases}
  \\
&
\begin{array}{llr}
a = -1 \ , &                                  b  =   6.1175316692890884760  \ , &     (\mbox{for}\,k_1)\ ,  \non  \\ 
a = 1.1137892269157833935\ ,  & b = 0.13093256295872348106 \ , &     (\mbox{for}\,k_2) \ , \non
\end{array}
 \non\\[10pt]
 I_5^{(5)} &=  r_{12}^2  + a r_{34}^2 + b r_{25}^2 +  r_{15}^2 =
 \begin{cases}
  -0.93759132182895716222 \ (\mbox{for}\,k_1)\ , \\
   \phantom{+}0.34530207099754785980 \ (\mbox{for}\,k_2)\ , \\
\end{cases}
\\
&
\begin{array}{llr}
a = -5.1175316692890884752 \ ,  &    b  =  1   \ ,    &  (\mbox{for}\,k_1)  \ ,  \non \\
a = 0.86906743704127652398\ ,  &    b  = -1.1137892269157833939 \ ,  & (\mbox{for}\,k_2)\ ,
\end{array}
\non \\[10pt]
 I_{6}^{(5)} &=    a r_{12}^2  + b r_{14}^2 + c r_{24}^2  + r_{35}^2
\begin{cases}
-3.8312045182313851648     \ (\mbox{for}\,k_1)\ , \\
 \phantom{+}9.5199479346308316909 \ (\mbox{for}\,k_2)\ , \\
\end{cases}
\\
&
\begin{array}{lllr}
 a = -5.9221249645292297285 \ , &        b   = -0.19540670475985874632\ ,  & c = b\ ,  &   (\mbox{for}\,k_1) \ , 
\non \\
 a = 1.1506586915791953876 \ ,  &  b   = a  \ , & c   =a \ , &   (\mbox{for}\,k_2)\ ,
\non
\end{array}
 \end{align}
\endwidetext
In total we have found 15 constants of motion. It can be checked by direct calculation
that {\it all} 15 constants correspond to the Liouville integrals along the algebraic Lemniscate (\ref{lemn}), having vanishing Poisson brackets with the Hamiltonian (\ref{fivebpotential}-\ref{V25b}).
Therefore,  the algebraic lemniscate is a particularly maximally superintegrable
trajectory for 5-bodies of equal mass moving choreographically. This fact represents
another example which supports the T-conjecture\cite{Turbiner:2013}.

It is worth noting that the potential function $V$ consists of two types of pairwise potentials, containing the logarithmic term or not.
For example, for $k_1$ case, the first type is represented by the
potential
\begin{equation}
\label{Vr125b}
  {\cal V}(r_{12}) \equiv \left\{  \alpha_1 {\log  r_{12}^2}  - \beta_1  r_{12}^2\right\}
\end{equation}
the nearest neighbors interaction, while the second type is represented by the potential
\begin{equation}
\label{Vr135b}
  {\cal V}(r_{13}) \equiv \left\{    - \beta_1  r_{13}^2\right\}\ ,
\end{equation}
the next-to-nearest neighbors interaction. For both potentials, the motion is bounded,
{\it i.e.} there exist a finite domain for
$$r_{12} \in  [r_{12}^{\rm min}, r_{12}^{\rm max}] = [0.6867\ldots,  1.1087\ldots]\ ,$$ lying in the domain of attraction of the potential (\ref{Vr125b}), and also a finite domain for
$$r_{13} \in [r_{13}^{\rm min}, r_{13}^{\rm max}]=[0.3841\ldots,1.8891\ldots]\ ,$$
although the potential (\ref{Vr135b}) is repulsive.
It is worth emphasizing that the parameter $\alpha_1$ in front of the logarithmic part of the potential (\ref{Vr125b}) remains the same as for the 3 body case (see  (\ref{HF})),
but $\beta_1$ in front of the repulsive harmonic oscillator interaction is smaller than one in the 3 body case.

As for $k_2$ case the pairwise potentials are of the types
$${\cal V}(r_{13}) \equiv \left\{  \alpha_2 {\log  r_{13}^2}  - \beta_2  r_{13}^2\right\}
\ ,$$ where
$$r_{13} \in [r_{13}^{\rm min}, r_{13}^{\rm max}] =  [0.6671\ldots,  1.995\ldots]\ ,$$
and
$${\cal V}(r_{12}) \equiv \left\{    - \beta_2  r_{12}^2\right\}\ ,$$
with
$$r_{12} \in [r_{12}^{\rm min}, r_{12}^{\rm max}] = [0.2663\ldots,  1.7913\ldots]\ .$$

Formally, the problem of 5 body choreographic motion on the algebraic lemniscate can be posed as a solution of the system of coupled Newton equations for the potential   (\ref{fivebpotential}). The initial conditions can be defined at the moment, say $t=0$, where each 3 bodies are situated on a straight line (called the Euler line), thus, we have two Euler lines evidently intersecting at the origin. In this case the convex hull of the set of bodies on the plane is a quadrangle with one body situated at an interior of it, at the origin). However, in general, the 5 body motion on the plane forms a degenerate (planar) pentagon (convex hull). In four (and higher) dimensional space, 5 bodies define a regular non-degenerate pentahedron which is characterized by 10 edges (relative distances). When the motion is projected to the plane, seven edges {\it only} are
independent. There must exist three constraints. One constraint is evident: the volume
of the pentahedron should vanish. It corresponds to degeneration to a 3-dimensional space. What are the other constraints? Answer to this question remains unknown to the present authors as well as how to approach to it.

\section{\it (C)\ 7-bodies choreography on the algebraic Lemniscate}

\begin{figure}
\subfloat[]{\includegraphics[width = 1.7in]{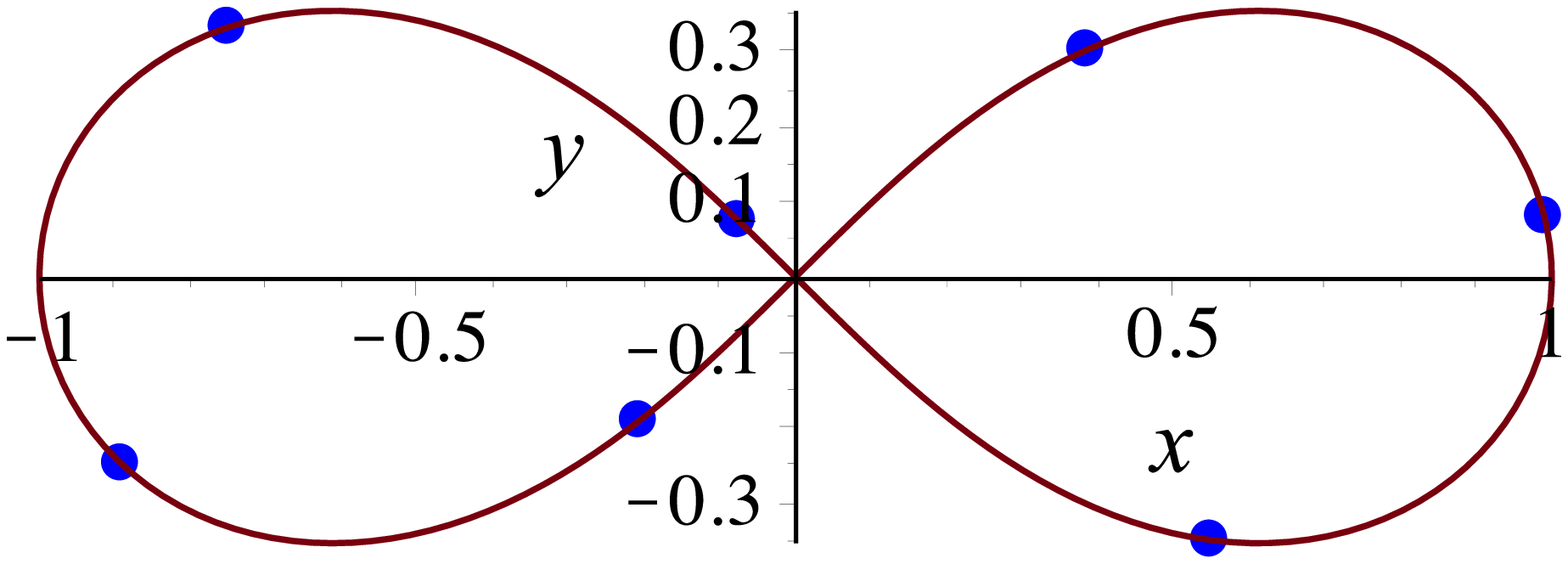}}\,
\subfloat[]{\includegraphics[width = 1.7in]{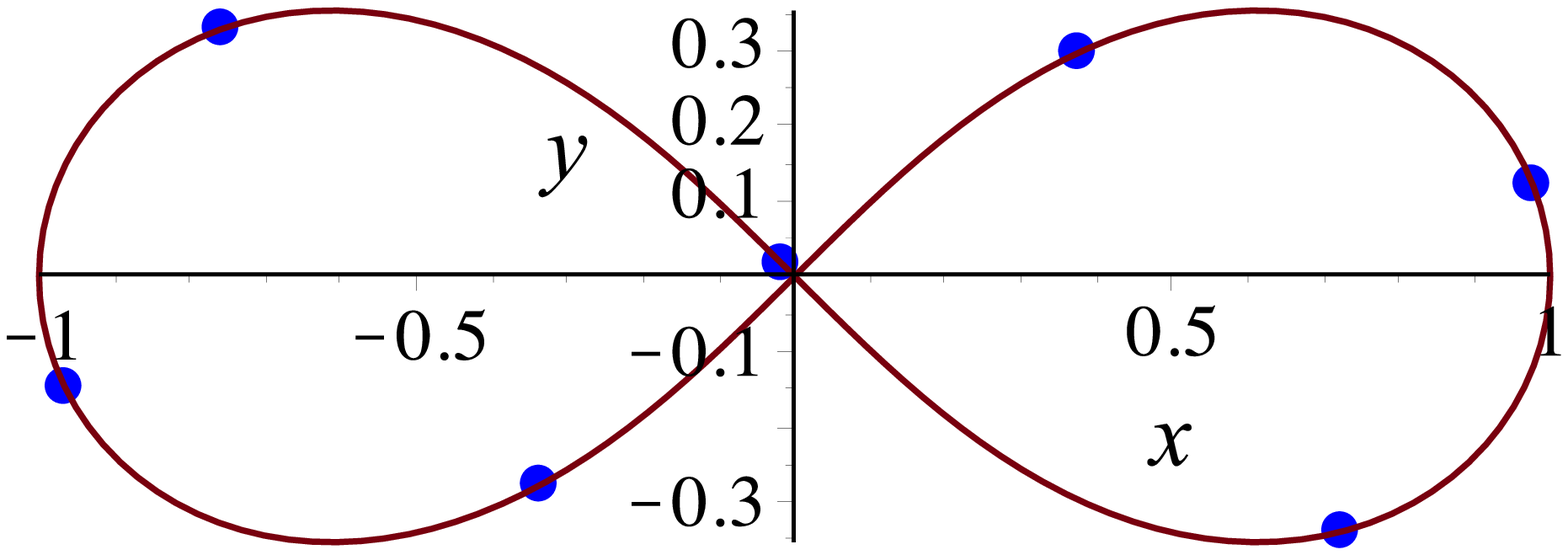}}\,
\subfloat[]{\includegraphics[width = 1.7in]{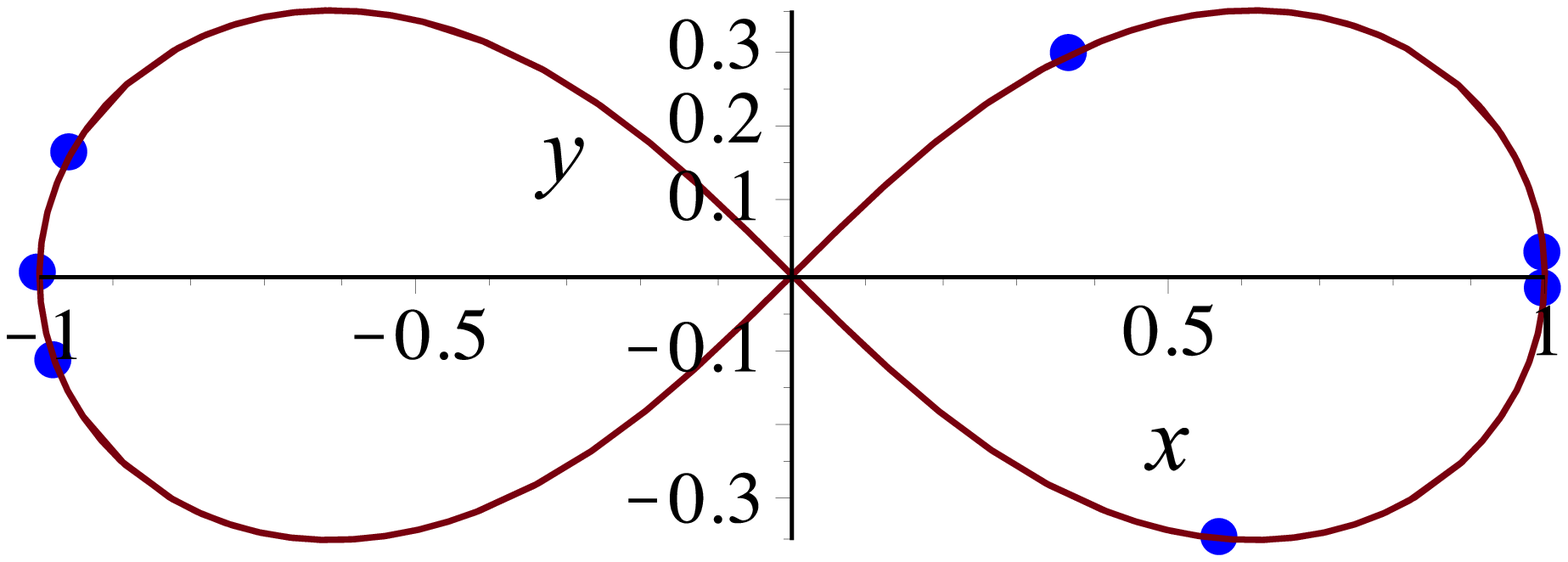}}\,
\caption{\label{fig8} Snapshots of seven equal masses moving on the algebraic Lemniscate by Bernoulli (\ref{lemn}) for $c=1$, (a) for $k_1$,  (b) for $k_2$ and (c) for $k_3$
(see text).}
\end{figure}

Three different choreographies of seven equal masses moving on a common
algebraic Lemniscate with fixed center-of-mass were found by Fujiwara
{\it et al.} in \cite{Fujiwara:2004} (see Fig.\ref{fig8}). Making analysis
one can show that these choreographies are three solutions of the system of fourteen
coupled Newton equations. They correspond to pairwise gravitational
force (in logarithmic term of the potential) among
(i) {\it nearest neighbors} (ii) {\it next-to-nearest neighbors}, and
(iii) next-to-next-nearest neighbors, plus a pairwise repulsive harmonic oscillator potential among all particles  (see below). For the choreography of seven bodies
moving on the algebraic lemniscate in order to be  maximally particularly superintegrable,
there should exist $23$ independent constants of motion. We are able to find some of them,
they are presented below.

We define a seven body choreography on the Lemniscate by placing
seven equal mass bodies on the curve (\ref{lemn}) with equal time-delay $\tau/7$. The
position of each particle is given by the plane vectors
\begin{align*}
  {\mathbf x}_1(t) &\ = ( x(t{ -3{\tau}/7}) \ ,  y(t-3{\tau}/7) ) \ , \\
  {\mathbf x}_2(t) &\ = ( x(t{ -2{\tau}/7}) \ ,  y(t-2{\tau}/7) ) \ , \\
  {\mathbf x}_3(t) &\ = ( x(t{ -{\tau}/7})  \ ,  y(t-{\tau}/7) )  \ , \\
  {\mathbf x}_4(t) &\ = ( x(t)\ ,\ y(t) )  \ ,                        \\
  {\mathbf x}_5(t) &\ = ( x(t{ +{\tau}/7})  \ ,  y(t+{\tau}/7) )  \ , \\
  {\mathbf x}_6(t) &\ = ( x(t{ +2{\tau}/7}) \ ,  y(t+2{\tau}/7) ) \ , \\
  {\mathbf x}_7(t) &\ = ( x(t{ +3{\tau}/7}) \ ,  y(t+3{\tau}/7) ) \ ,
\end{align*}
where $x(t)$ and $y(t)$ are given by the algebraic Lemniscate's parametrization
(\ref{param}), and ${\tau}=4K(k)$ is the period of the motion. It was
found that such period is defined by the condition that the center of
mass of the system will remain fixed, {\it i.e.}
\[
 \resizebox{.95\hsize}{!}{$
  {\mathbf X}_{ \sc   CM}\ =
  {\mathbf x}_1(t) + {\mathbf x}_2(t) + {\mathbf x}_3(t) + {\mathbf x}_4(t)
  + {\mathbf x}_5(t) + {\mathbf x}_6(t)+ {\mathbf x}_7(t)=0$}\ ,
\]
\[
 \resizebox{.95\hsize}{!}{$
  {\mathbf V}_{ \sc   CM}\ =
  {\mathbf v}_1(t) + {\mathbf v}_2(t) + {\mathbf v}_3(t) + {\mathbf v}_4(t)
  + {\mathbf v}_5(t) + {\mathbf v}_6(t)+ {\mathbf v}_7(t)=0$}\ .
\]
This condition is satisfied only for three values of the elliptic modulus $k$.
These values appear as a solution of the following equation derived in \cite{Fujiwara:2004},
\begin{equation}
\label{CMdncondition}
 \mathrm{dn}(z_0(k),k)\ =\ \frac{1}{\sqrt{2}}\ ,
\end{equation}
cf.(\ref{3body-k}),(\ref{5body-k}), where the values of the argument $z_0$ with the corresponding solutions for the square of the elliptic modulus, are
\begin{align}
  z_0 &=\frac{9K}{7}  , \quad k_1^2=0.574 569 280 934 588 654 057 271 2 \ , \non \\
  z_0 &=\frac{11K}{7} , \quad k_2^2=0.830 609 000 670 624 071 081 777 9 \ , \label{ks7b}\\
  z_0 &=\frac{13K}{7} , \quad k_3^2=0.999 930 000 538 037 277 285 982 8 \ . \non
\end{align}
The argument in (\ref{CMdncondition}) is given by
$$z_0= K + m\, \frac{\delta z_0}{2}\ , m=1,2,3\ ,$$
where $\delta z_0 = T/7= 4K/7$ is the time delay between the neighbouring bodies.
The corresponding periods, see (\ref{period}), are
  \[
 \tau(k)\ = \begin{cases}
   7.69200191400285631133\  \ \mbox{for } k_1 \ ,  \\
   9.33221968323490708873\  \ \mbox{for } k_2 \ ,  \\
  24.67958534628126681073\  \ \mbox{for } k_3 \ ,  \\
\end{cases}
\]
cf.(\ref{period-3}).

\begin{figure}
\subfloat[]{\includegraphics[width = 1.7in]{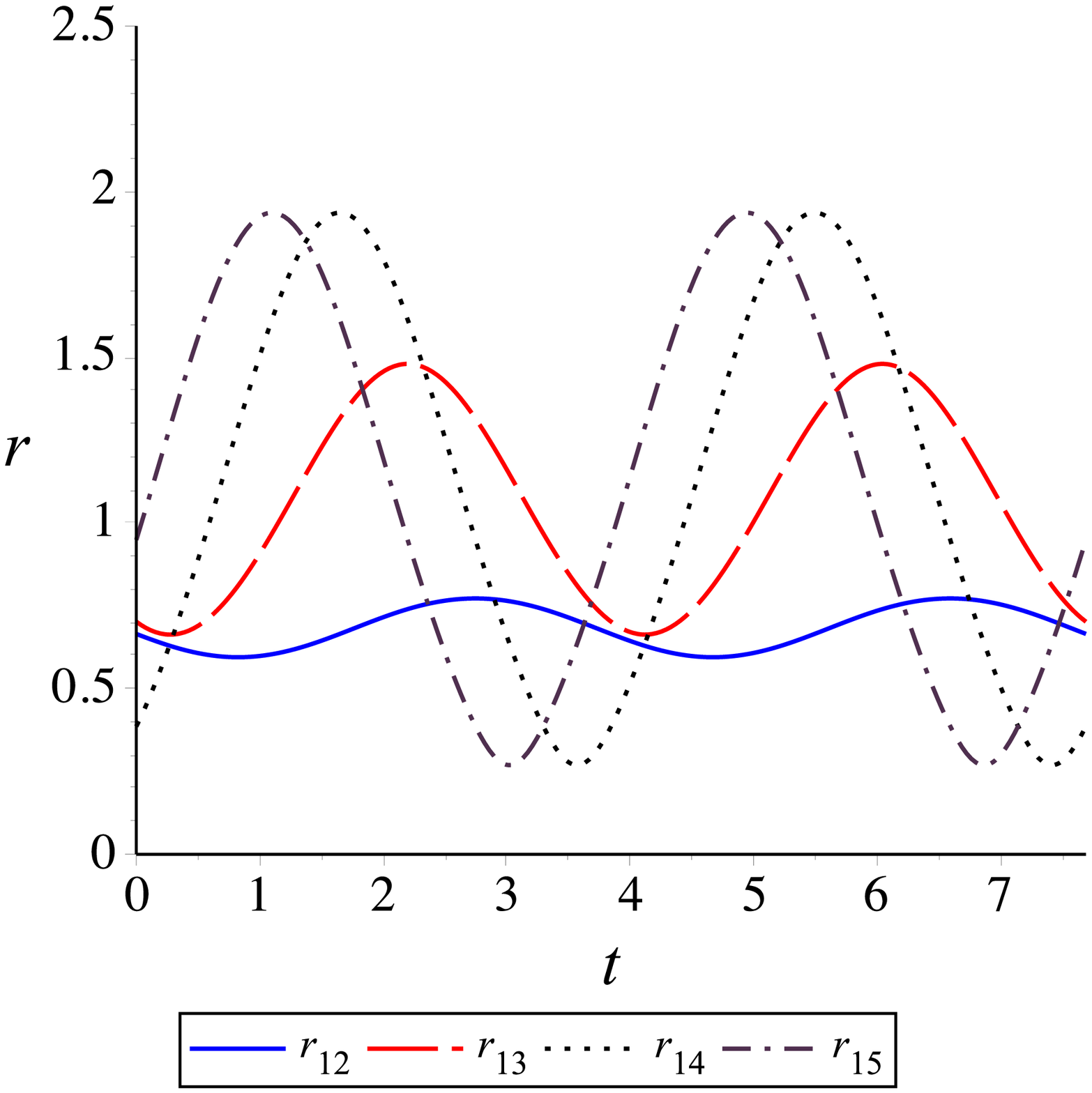}}\,
\subfloat[]{\includegraphics[width = 1.7in]{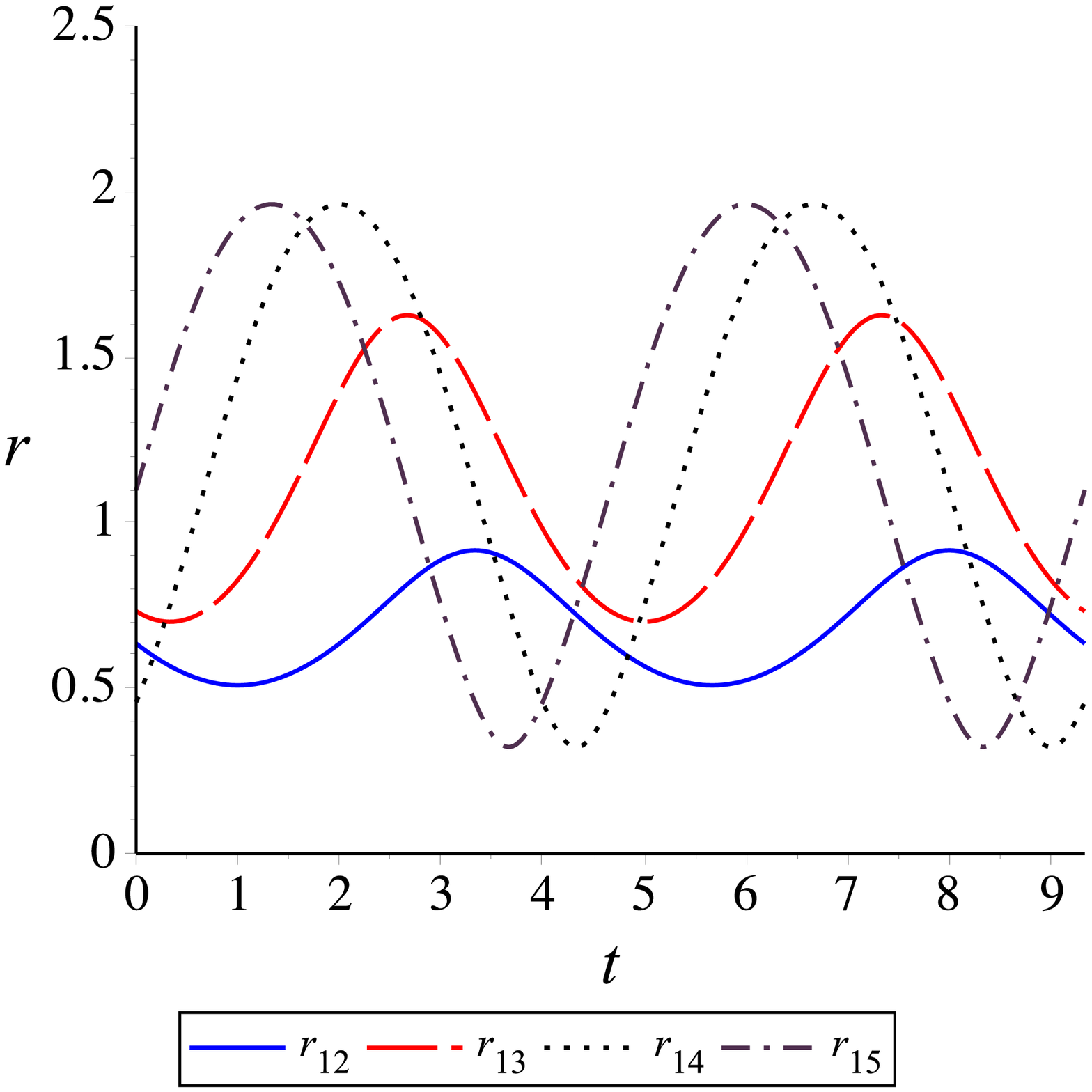}}\,
\subfloat[]{\includegraphics[width = 1.7in]{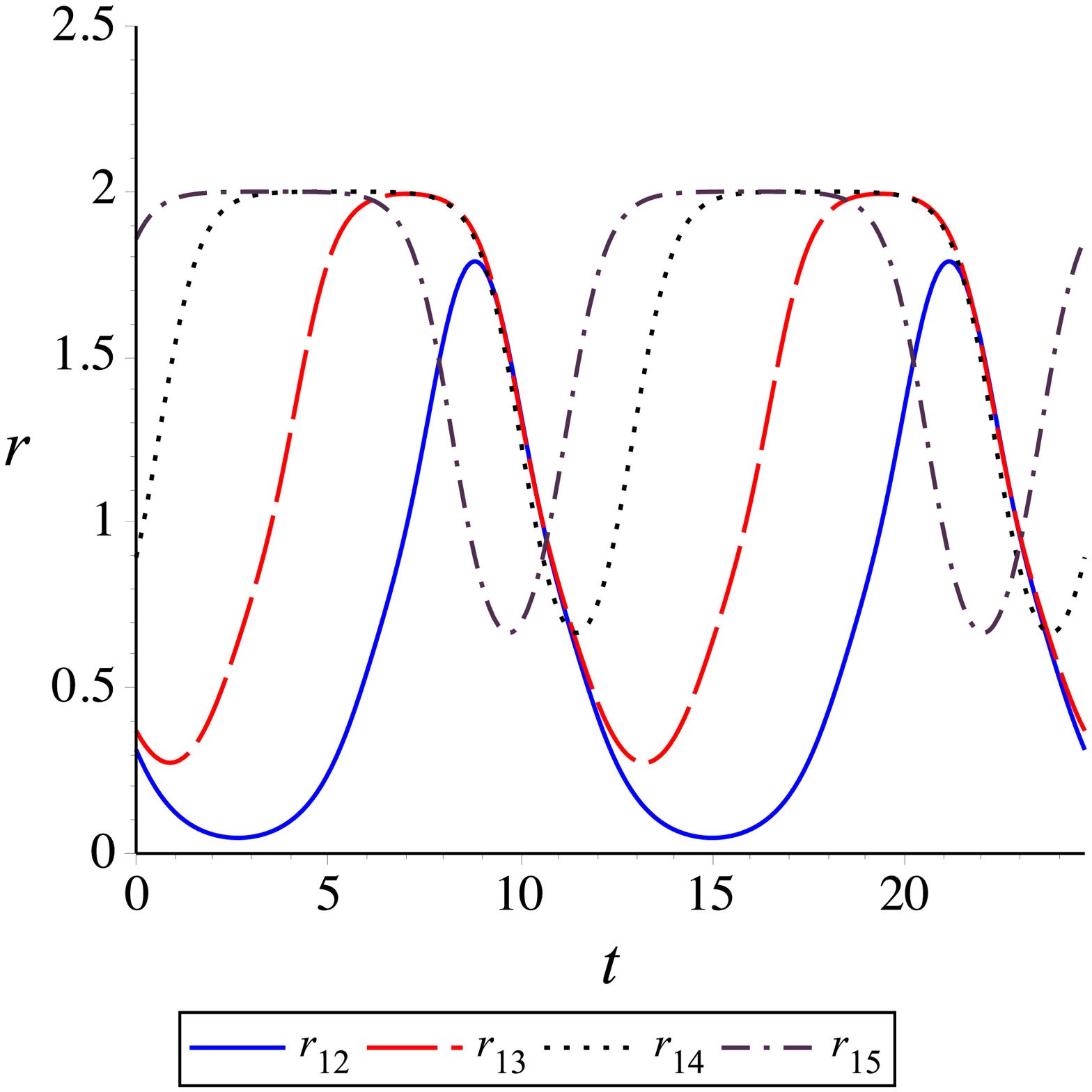}}\,
\caption{\label{fig9} Relative distances of seven equal masses moving on the algebraic Lemniscate by Bernoulli (\ref{lemn}) for $c=1$, (a) for $k_1$,  (b) for $k_2$ and
(c) for $k_3$ (see text).}
\end{figure}

\begin{figure}
\subfloat[]{\includegraphics[width = 1.7in]{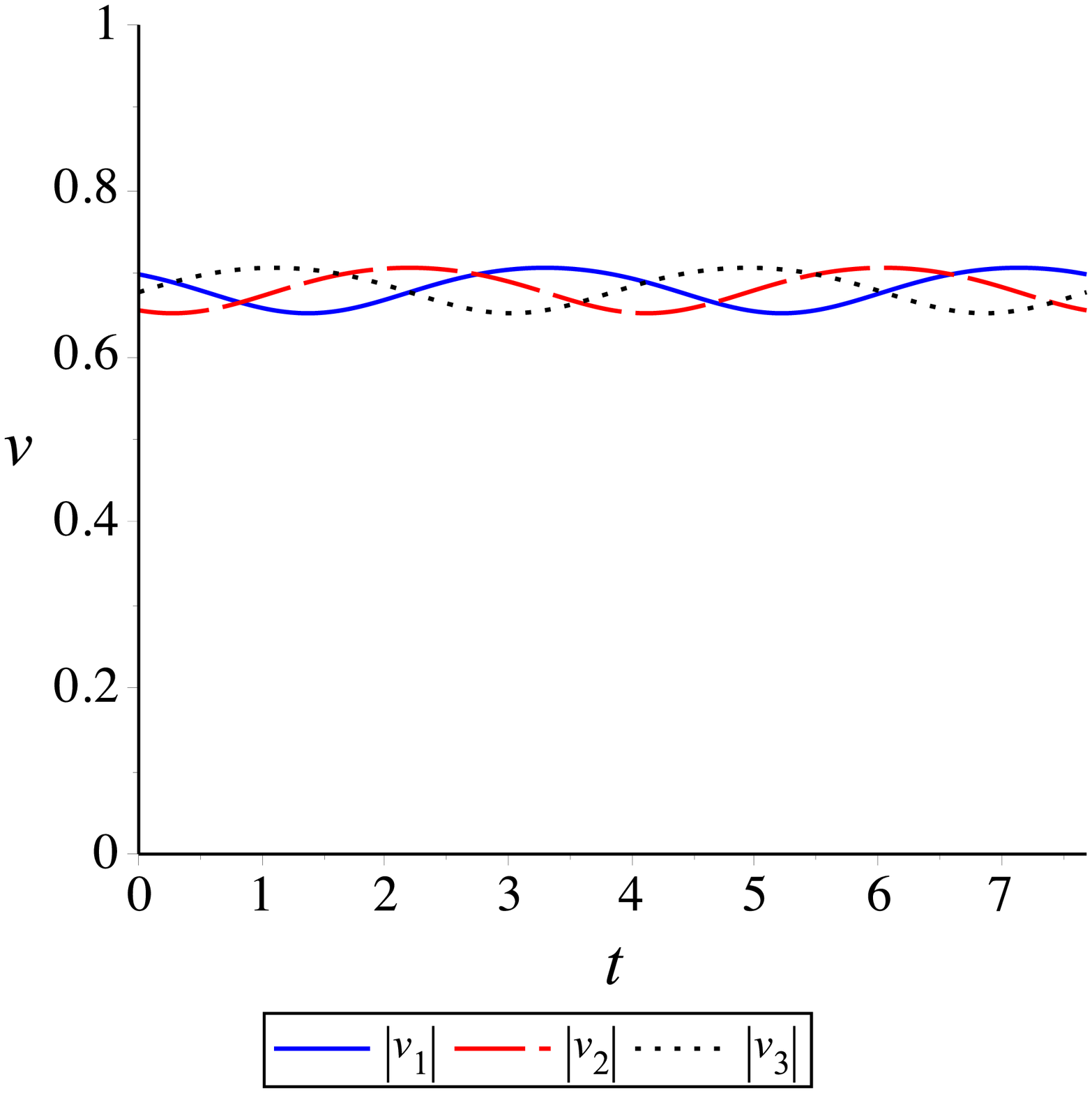}}\,
\subfloat[]{\includegraphics[width = 1.7in]{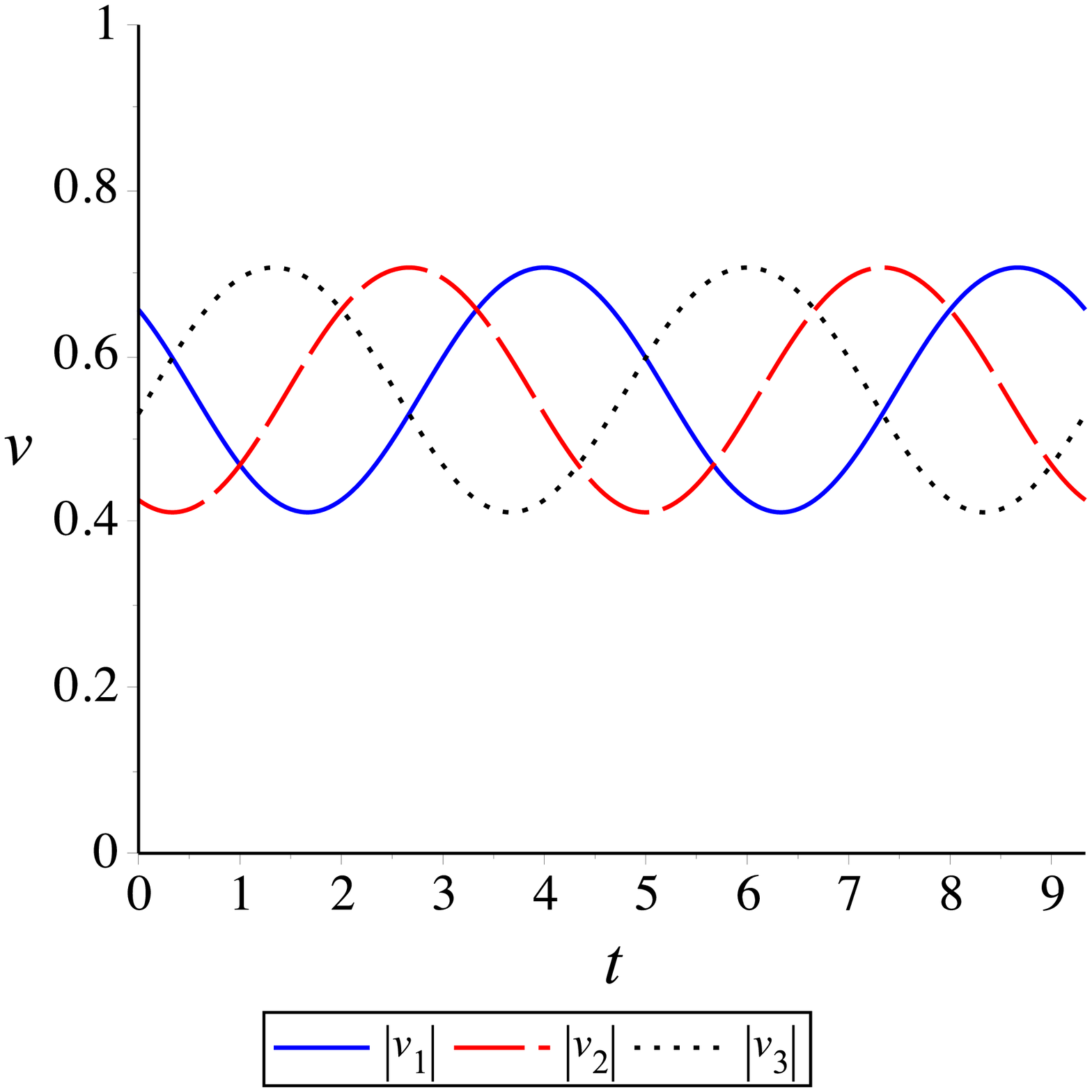}}\,
\subfloat[]{\includegraphics[width = 1.7in]{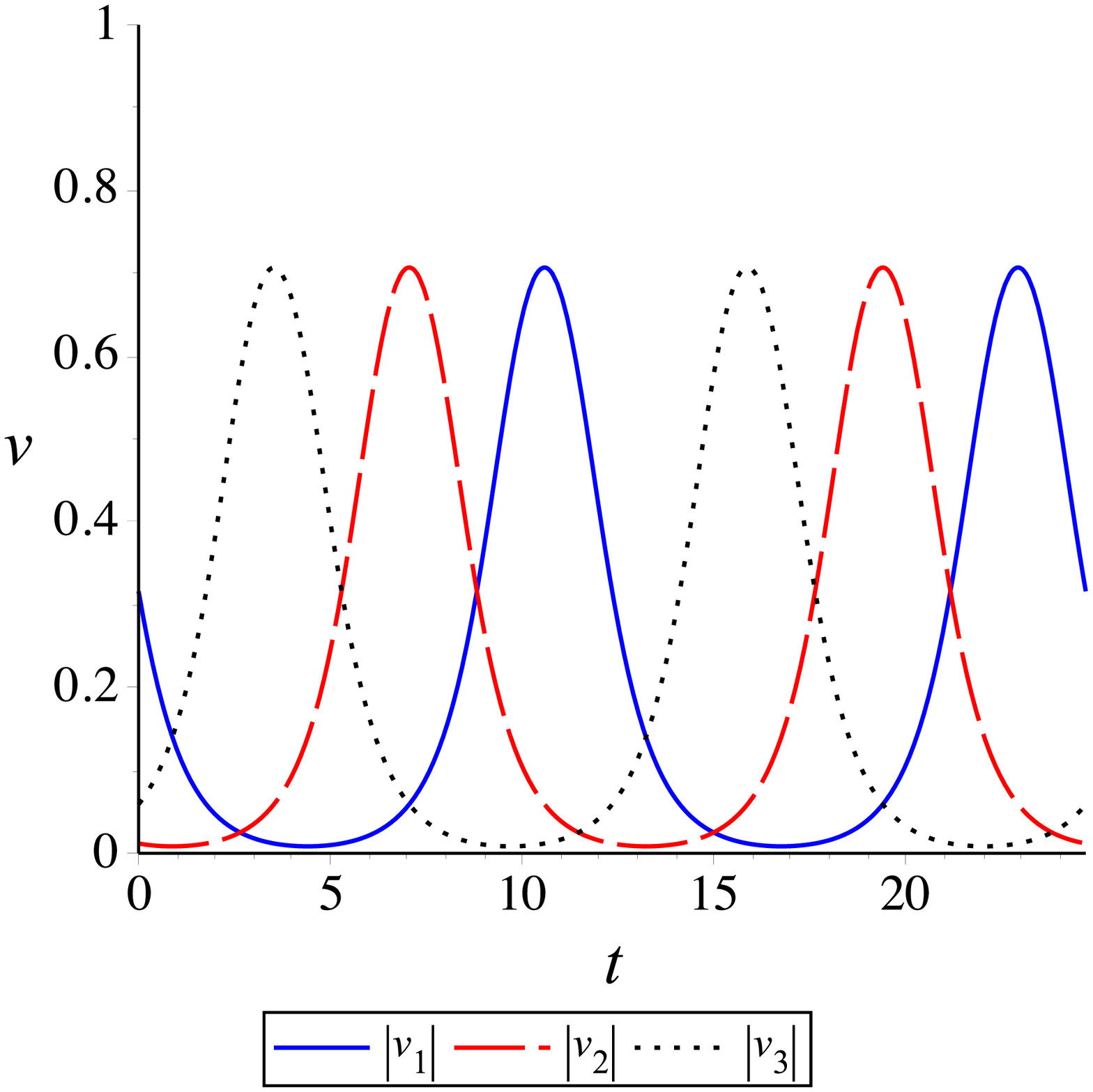}}\,
\caption{\label{fig10} Velocities $|{\mathbf v}_{1,2,3}|$ of the seven equal masses moving on the algebraic Lemniscate by Bernoulli (\ref{lemn}) for $c=1$,
(a) for $k_1$, (b) for $k_2$ and (c) for $k_3$ (see text). For all velocities for all three cases the maximal velocity $v^{\rm max}\, =\, 1/\sqrt{2}$ is the same, it corresponds to a moment when a body passes through the crossing point of the lemniscate  ${\mathbf x}=0$ (see relation (\ref{xvrelation})).}
\end{figure}

Snapshots of the motion of these three choreographies are shown
in Fig.\ref{fig8}, and the time evolution of some relative distances and absolute values of the velocities
in Figs.\ref{fig9},\ref{fig10}, in general, their time evolution is periodic with half of the total period. In particular, from these figures one can see that
for the smallest value of the elliptic modulus $k=k_1$ the motion becomes more uniform as the oscillation of the
absolute value of the velocities $|{\mathbf v}_i|\,,i=1,2\ldots 7$ occurs in a rather narrow domain, while
for the largest value $k=k_3$ of the elliptic module, the oscillation of each velocity has larger amplitude and eventually two bodies approach to each other closely both having very small velocity (see Figs.\ref{fig9}(c), \ref{fig10}(c).

Apart from the total energy $E={\cal T} + {\cal V}$ and the total angular momentum
$L=\sum_{i=1}^7 {\mathbf x}_i \times {\mathbf v}_i = 0$, which are global integrals of motion, there exist three special, velocity-independent
constants of motion along the trajectory, different for each $k$:

\widetext

\begin{align}
I_1^{(7)} &=
  \begin{cases}
 r_{12}^2 r_{23}^2 r_{34}^2 r_{45}^2 r_{56}^2 r_{67}^2 r_{17}^2= 0.26362178303408707110 \  (\mbox{for}\,k_1)\ ,
 \\[3pt]
 r_{13}^2 r_{35}^2  r_{57}^2  r_{27}^2  r_{24}^2 r_{46}^2 r_{16}^2 = 2.6489374483809056078 \ (\mbox{for}\,k_2)\ ,
 \\[3pt]
 r_{14}^2 r_{47}^2 r_{37}^2  r_{36}^2 r_{26}^2 r_{25}^2 r_{15}^2 = 482.72504286636644935 \ (\mbox{for}\,k_3)\ ,
\end{cases}
\label{I17b}\\
 I_2^{(7)} &=
  \begin{cases}
  r_{12}^2 + r_{23}^2 + r_{34}^2 + r_{45}^2 + r_{56}^2 + r_{67}^2 + r_{17}^2=   3.3066909304564091707 \  (\mbox{for}\,k_1)\ ,
 \\[3pt]
  r_{13}^2 + r_{35}^2  + r_{57}^2  + r_{27}^2  + r_{24}^2 + r_{46}^2 + r_{16}^2 =  9.5236213317249233435 \ (\mbox{for}\,k_2)\ ,
 \\[3pt]
  r_{14}^2 + r_{47}^2 + r_{37}^2  + r_{36}^2 + r_{26}^2 + r_{25}^2 + r_{15}^2 =  20.460348437174532420 \ (\mbox{for}\,k_3)\ ,
  \end{cases}
\label{I27b}\\
 I_{HR}^{(7)}&= 7 \sum {\mathbf x}_i^2= \sum_{i<j} r_{ij}^2  = \begin{cases}
22.883408614201644590 \, (\mbox{for}  \, k_1)  \ , \\
25.662798924376851272 \, (\mbox{for}  \, k_2) \ ,  \\
39.103415650888148540 \, (\mbox{for}  \, k_3) \ , \\
\end{cases} \label{IHR7b}
\end{align}
\noindent
where we identify  $I_{HR}^{(7)}$ as the moment of inertia, or the hyperradius squared
in the space of relative distances, {\it i.e.} the sum of the squares of the 21 different relative distances among the bodies.

The kinetic energy is also a constant of motion
\begin{equation}
{T}=\frac{1}{2}\sum {\mathbf v}_i^2 = \begin{cases}
 1.6281143338790436808 \, (\mbox{for}  \, k_1) \ , \\
 1.1439748352286144920 \, (\mbox{for}  \, k_2) \ , \\
 0.35364495661517090077\, (\mbox{for}  \, k_3) \ .  \\
\end{cases}
\end{equation}
Therefore, we can assume that the potential, being a constant of motion, is made out of the above velocity independent constants. Moreover, if we request $V$ to be
composed of  pairwise interactions only we propose the following Ansatz:
\begin{equation}
\label{sevenpotential}
  {\cal V}= \alpha_7  \log {I_1^{(7)}} - \beta_7  I_{\rm HR}^{(7)}\ .
 \end{equation}
It was found  that, for each possible value of $k$, there exist $\alpha, \beta$ s.t.
${\cal V}$ satisfies the system of fourteen coupled Newton equations
\[
  \frac{d^2}{d t^2} {\mathbf x_i} (t)\   =\  -{\nabla_{\mathbf x_i} {\cal V}}\ , \quad  {\scriptstyle  i=1 \ldots 7} \ ,
\]
or, equivalently, twelve coupled Newton equations for the relative motions
in eleven independent variables $r_{ij}$. Explicitly, it is found that

\begin{equation}
\label{potential7b}
 {\cal V}=  \begin{cases}
 {  \alpha_1}{ \left\{ \log  r_{12}^2 + \log  r_{23}^2 +\log  r_{34}^2 +\log  r_{45}^2 +\log  r_{56}^2 +\log  r_{67}^2 +\log  r_{17}^2 \right\} }  \\
 {  \alpha_2}{ \left\{ \log  r_{13}^2 + \log  r_{35}^2 +\log  r_{57}^2 +\log  r_{27}^2 +\log  r_{24}^2 +\log  r_{46}^2 +\log  r_{16}^2 \right\} }  \\
 {  \alpha_3}{ \left\{ \log  r_{14}^2 + \log  r_{47}^2 +\log  r_{37}^2 +\log  r_{36}^2 +\log  r_{26}^2 +\log  r_{25}^2 +\log  r_{15}^2 \right\} }  \\
 \end{cases}
 -  {  \beta_{1,2,3}} \sum_{i<j} r_{ij}^2 \ ,
\end{equation}
with
\begin{align}
\label{potential7c}
 \alpha_1 &= \frac{1}{4}\ ,\ \beta_1 = 0.0053263772096134752826 \ (\mbox{for}\,k_1)  \ , \nonumber \\
 \alpha_2 &= \frac{1}{4}\ ,\ \beta_2 = 0.023614928619330290764 \ (\mbox{for}\,k_2) \ ,
 \\
 \alpha_3 &= \frac{1}{4}\ ,\ \beta_3 = 0.035709285752716948625 \ (\mbox{for}\,k_3)\ ,  \nonumber
\end{align}
satisfies the Newton equations.

In addition to the above-mentioned constants of motion ($E, {L}, I_1^{(7)}, I_2^{(7)}, I_{\rm HR}^{(7)},T$), where, in particular, the total energy $E$ takes values
\[
 E\ =\ {\cal T} + {\cal V}\ =\\
 \begin{cases}
0.16423442473755265532 \ \ (\mbox{for}\ k_1)\ ,
 \\
0.78148931963620224401 \ \ (\mbox{for}\ k_2)\ ,
\\
0.50215171876269155046 \ \ (\mbox{for}\ k_3)\ ,
 \end{cases}
\]
%
%
%
%
%
it can be shown that the following functions are also constants of motion on the algebraic lemniscate (\ref{lemn}):
 \begin{equation}
\tilde{\cal T}\ =\
{\mathbf v}_1^2{\mathbf v}_2^2{\mathbf v}_3^2{\mathbf v}_4^2
{\mathbf v}_5^2{\mathbf v}_6^2{\mathbf v}_7^2 =
\begin{cases}
 0.00466005751814023926 \ (\mbox{for}\,k_1)\ , \\
 0.00024297642722229551  \ (\mbox{for}\,k_2)\ , \\
 1.188916715465945825\times 10^{-15} \ (\mbox{for}\,k_3)\ ,
 \end{cases}
 \end{equation}
 %
 and
\begin{align}
J_i(k_{1,2,3})  &\ =\
{\mathbf v}_i^2 + \frac{1}{49}\left(k_{1,2,3}^2 - \frac{1}{2} \right) \non \times \\
&
\left( b^{(i)}_1 r_{12}^2+ b^{(i)}_2 r_{13}^2
          + b^{(i)}_3 r_{14}^2+ b^{(i)}_4 r_{15}^2 + b^{(i)}_5 r_{16}^2+ b^{(i)}_6 r_{17}^2
          + b^{(i)}_7 r_{23}^2+ b^{(i)}_8 r_{24}^2 + b^{(i)}_9 r_{25}^2+ b^{(i)}_{10} r_{26}^2 + b^{(i)}_{11} r_{27}^2
          \right.\non  \\ & \left.
          + b^{(i)}_{12} r_{34}^2 + b^{(i)}_{13} r_{35}^2+ b^{(i)}_{14} r_{36}^2 + b^{(i)}_{15} r_{37}^2
          + b^{(i)}_{16} r_{45}^2  + b^{(i)}_{17} r_{46}^2  + b^{(i)}_{18} r_{47}^2  + b^{(i)}_{19} r_{56}^2  + b^{(i)}_{20} r_{57}^2 + b^{(i)}_{21} r_{67}^2     \right)\ =\ \frac{1}{2}\ , \\
          &\ i=1\ldots 7\ , \non
\end{align}
\endwidetext
\noindent
for certain coefficients $b_{1,2\ldots 21}^{(i)}$ with the property $\sum_{i=1}^7 b _{1,2\ldots 21}^{(i)}=7$. These variables   are then constrained to satisfy the relation
\[
\sum^7 J_i(k_0) = 2\ {T} + \frac{1}{7} (k_{1,2,3}^2-\frac{1}{2})   I_{\rm HR}^{(7)}\ .
\]
Note that in a similar way as for 3-body and 5-body choreographies the quantities ${\tilde{\cal T}}$ and $I_1^{(7)}$ play the role of dual quantities $r^2_{i,i+1} \lrar v^2_i$, as well as for $T$ and $I_{\rm HR}^{(7)}$.   The quantity ${\tilde{\cal T}}$, the product of velocities squared, takes a very small value for the choreography with elliptic modulus close to one, i.e. $k_3$.
It indicates that at some moment of evolution two bodies have a very small velocity (see Fig.\ref{fig10} (c)). This occurs at the end points of the lobes of the lemniscate, where two bodies approach to each other.

Thus far, we have in total 12 constants of motion out of 23. So, for the algebraic Lemniscate to be a maximally particularly superintegrable trajectory of the  7-body choreography, and verify the T-conjecture, we need to find 11  constants more.  What are the missing particular constants of motion?
A natural hint suggests to consider, for instance, polynomials in relative distances only
with non-integer coefficients, as it was done for the 5-body case. This procedure is lengthy and it will not be presented here.

As it occurs for the five body case, the pairwise potential function consists of two types of pairwise potentials. For example, for $k_1$, one type is represented by the potential
 \begin{equation}
 \label{Vr127b}
  {\cal V}(r_{12}) \equiv \left\{  \alpha_1 {\log  r_{12}^2}  - \beta_1  r_{12}^2\right\} ,
 \end{equation}
the nearest neighbors interaction, while the second type is represented by the potentials
 \begin{equation}
 \label{Vr137b}
  {\cal V}(r_{13}) \equiv \left\{    - \beta_1  r_{13}^2\right\},  \quad {\cal V}(r_{14}) \equiv \left\{    - \beta_1  r_{14}^2\right\},
 \end{equation}
the next-to-nearest neighbors and next-to-next nearest neighbors interactions.
In all three cases, the motion is bounded, {\it i.e.} there exist a finite domain for
%
$$
 r_{12} \in [r_{12}^{\rm min}, r_{12}^{\rm max}] = [0.5949 \ldots ,  0.7723\ldots]\ ,
$$ lying in the attractive sector of the potential (\ref{Vr127b}), and a finite domain for
$$
 r_{13} \in [r_{13}^{\rm min}, r_{13}^{\rm max}] = [0.6632\ldots,1.4802\ldots]\ ,
$$
and
$$
 r_{14} \in [r_{14}^{\rm min}, r_{14}^{\rm max}]=[0.2692\ldots,1.9365\ldots]\ ,
$$
despite the fact that the potentials (\ref{Vr137b}) are purely repulsive. It is worth to note that the strength $\alpha_1$ of the logarithmic part of the potential remains the
same as that in the three and five body cases. It seems to be a quantity independent of the number of bodies. On the other side, the strength $\beta_1$ of the repulsive harmonic interaction is much weaker than that in the five and three body cases.

As for $k_2$ case the pairwise potentials are of the types
$${\cal V}(r_{13}) \equiv \left\{  \alpha_2 {\log  r_{13}^2}  - \beta_2  r_{13}^2\right\}
\ ,$$ where
$$r_{13} \in [r_{13}^{\rm min}, r_{13}^{\rm max}] = [0.7002\ldots,  1.6263\ldots]\ ,$$
and
$${\cal V}(r_{12}) \equiv \left\{    - \beta_2  r_{12}^2\right\}\ , \quad {\cal V}(r_{14}) \equiv \left\{    - \beta_2  r_{14}^2\right\}\ ,  $$
with
$$r_{12} \in [r_{12}^{\rm min}, r_{12}^{\rm max}] = [0.5082\ldots,  0.9157\ldots]\ ,$$
$$r_{14} \in [r_{14}^{\rm min}, r_{14}^{\rm max}] = [ 0.3219\ldots,  1.9618 \ldots]\ .$$
In this case, the strength $\beta_2$ of the repulsive harmonic interaction is also weaker than that in the three body case, but stronger than the corresponding next-to-nearest neighbors interaction for the five body case.
%

In a similar way for $k_3$ case the pairwise potentials are of the types
$${\cal V}(r_{14}) \equiv \left\{  \alpha_3 {\log  r_{14}^2}  - \beta_3  r_{14}^2\right\}
\ ,$$ where
$$r_{14} \in [r_{14}^{\rm min}, r_{14}^{\rm max}]\ =\ [0.6666\ldots,  1.9998\ldots]\ ,$$
and
$${\cal V}(r_{12}) \equiv \left\{ - \beta_3  r_{12}^2\right\}\ , \quad {\cal V}(r_{13}) \equiv \left\{ - \beta_3  r_{13}^2\right\}\ , $$
with
$$r_{12} \in [r_{12}^{\rm min}, r_{12}^{\rm max}] = [0.0472\ldots,  1.7889\ldots]\ ,$$
$$r_{13} \in [r_{13}^{\rm min}, r_{13}^{\rm max}] = [ 0.2742\ldots, 1.9931\ldots]\ .$$
In this case, the strength $\beta_3$ of the repulsive harmonic interaction is also weaker than that in the three body case.

\section{\it (D)\ Choreographies of $(2n+1)$ bodies on the algebraic Lemniscate}

Now let us consider a choreography of $2n+1$ ($n \in  {\mathbb N}$) bodies on the algebraic lemniscate, which are defined by the time dependent position vectors:
\begin{align}
  x_{j} &\ =\ x\left(t { -(n+1-j)\frac{\tau}{2n+1}}  \right)\ , \\
  y_{j} &\ =\ y\left(t { -(n+1-j)\frac{\tau}{2n+1}}  \right)\ , \non \\
 j&=1\ldots 2n+1 \non\ .
\end{align}
It corresponds to the positions of $2n+1$ bodies situated along the algebraic Lemniscate    with equal time-delays $\tau/(2n+1)$. The condition for a fixing the center-of-mass
\[
  {\mathbf X}_{\rm CM}(t) =
  {\mathbf x}_1 + {\mathbf x}_2 +  \ldots + {\mathbf x}_{2n+1}=0\ ,
\]
\[
  {\mathbf V}_{\rm CM}(t) =
  {\mathbf v}_1 + {\mathbf v}_2 +  \ldots + {\mathbf v}_{2n+1}=0\ ,
\]
is satisfied by some number of solutions which obey the Fujiwara et al,'s equation (\ref{CMdncondition}) \cite{Fujiwara:2004},
where  the $z_0$ values are chosen requiring
\[
 z_0\ =\ K + m \, \frac{\delta z_0}{2} < 2K \ , \quad m = 1,2,3 \ldots\ m_{max} \ ,
\]
where  $\delta z_0 = \tau/(2n+1)= 4K/(2n+1)$  is the time delay between the bodies. Equivalently,
\begin{equation}
\label{nz0}
  z_0 = \left(1 + \frac{2m}{2n+1}\right) \, K \ , \quad m=1,2\ldots n\ ,
\end{equation}
cf.(\ref{3body-k}),(\ref{5body-k}),(\ref{ks7b}).  With these values, the relation (\ref{CMdncondition}) yields the possible values of $k$ such that two poles of $x^{(+)}(t) = x(t) + i y(t)$ (two out of its four poles in the fundamental domain, with the same imaginary part) have the same time distance as the time delay between bodies. When the sum $f(t) = \sum x_i^{(+)}(t) =0$ is considered, the   individual poles (and residues) are cancelled out. Such a choice guarantees the conservation of the
center-of-mass, the angular momentum and the moment of inertia, since all these quantities share the same pole structure (see \cite{Fujiwara:2004} for details).

Thus, for $(2n+1)$-bodies on the algebraic Lemniscate there exist $n$ different values for the elliptic modulus: $k_m\ , \quad m=1\dots n$, of the equation (\ref{CMdncondition}) and correspondingly $n$ different choreographies, each one of them is characterized by its own period and total energy. As a result of analysis one can draw a conclusion that all $k_{1\ldots n}^2 \in [1/2, 1]$. Making ordering $1/2 < k_1^2 < k_2^2 < \ldots < k_n^2 < 1$ one can see that the period (\ref{period}) grows with $k^2$,  since the elliptic integral $K(k^2)$ is a monotonous growing function of $k^2$. Minimal period always corresponds to $k_{1}^2$.  The  analysis which was done for the cases $(2n+1)=3,5,7,\ldots 21$ indicates that all $n$ choreographies are the solutions of the system of $(4n)$ coupled Newton equations with a potential which is a superposition of logarithmic term and inverted harmonic oscillator potential (see below).  Each choreography is characterized by its total energy $E$: for a given number of bodies $(2n+1)$ the minimal total energy $E_{\rm min}$ always corresponds to $k_1$, the minimal value of $k$, see Fig.11 (with the only exception of the five body case $n=2$). As the function of $n$ the minimal energy $E(k_1^2)$ grows for $n=1,2$ and then starts to decrease. For 9-body choreography (at $n=4$) the energy becomes negative and then tends to minus infinity as $(-n \ln n)$ (see below), when $n\to \infty$, see Fig.\ref{fig11}. Note that when the number of bodies grows the minimal value $k_{min}=k_1(n)$ decreases monotonously approaching to $k^2=1/2$, see Fig.\ref{fig12}. The period $\tau(k_1^2)$ is a monotonously decreasing function of $n$, see Fig.\ref{fig13}: it approaches asymptotically to a constant as $n$ grows.

\begin{figure}
 \includegraphics[width = 2.5in]{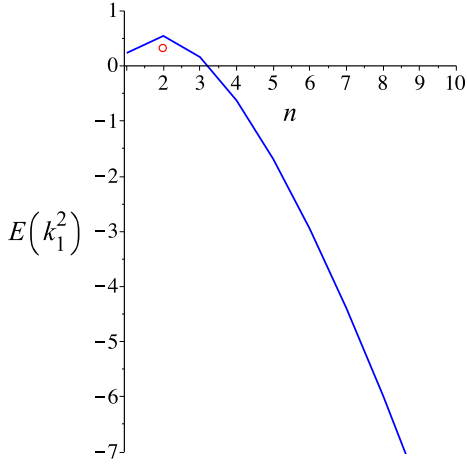}
\caption{\label{fig11}  Energy $E(k_1^2)(n)$ {\it vs} $n$  for the $(2n+1)$-body choreography with minimal value $k_{min}=k_1(n)$. For $n=2$ (5-body exceptional case)
$k_{min}=k_2(2)$ marked by red (see text). }
\end{figure}

\begin{figure}
 \includegraphics[width = 2.5in]{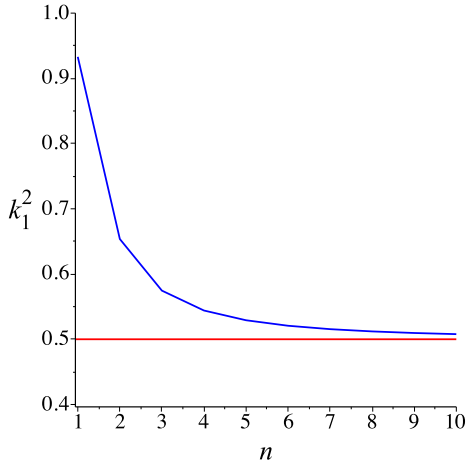}
\caption{\label{fig12}  Minimal elliptic modulus $k_1^2$ {\it versus} $n$  for the $(2n+1)$-body choreography with $n=1,2,\ldots , 10$.
The horizontal line corresponds to the limit $\lim_{n \to \infty} k_1^2  = \frac{1}{2}$  (see text).}
\end{figure}

\begin{figure}
 \includegraphics[width = 2.5in]{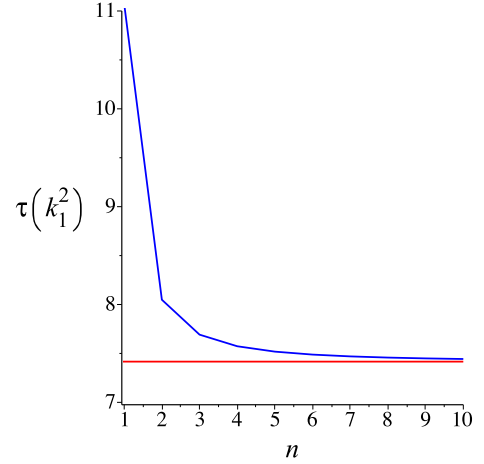}
\caption{\label{fig13}  Period $\tau(k_1^2)$ {\it vs} $n$  for  the $(2n+1)$-body choreography { for $n=1,2,\ldots , 10$} with minimal value $k_{min}=k_1$. The horizontal line corresponds to the asymptotic value $\lim_{n \to \infty} \tau(k_{1}^2) = \frac{\Gamma\left(\frac{1}{4}\right)^2}{\sqrt{\pi}}$ (see text).}
\end{figure}

For each $(2n+1)$-body choreography, additionally to the total energy $E$
and the total angular momentum $L=\sum_{i=1}^{(2n+1)} ({\mathbf x}_i \times
{\mathbf v}_i)\ =\ 0$, which are global conserved integrals, the moment of inertia
(the sum of all $(2n+1)n$ relative distances squared, {\it i.e.} the square of the hyper-radius) $I_{\rm HR}^{(2n+1)}$ is a constant of motion for the evolution:
\begin{equation*}
  I_{\rm HR}^{(2n+1)}\ =\ (2n+1) \sum_{i=1}^{(2n+1)} {\mathbf x}_i^2\ =\
   \sum_{i<j} r_{ij}^2  = \mbox{const}\ .\
\end{equation*}
for all $k_m, m=1,2, \ldots , n$ solutions, the kinetic energy is also a constant of motion. This is a consequence of the relation (\ref{xvrelation}),  {\it i.e.} the conservation of the moment of inertia implies that the kinetic energy is also conserved:
\begin{equation}
  {\cal T}\ =\ \frac{1}{2} \sum_{i=1}^{(2n+1)} {\mathbf v}_i^2  =  \mbox{const}\  .
\end{equation}

The general scheme suggests to guess that the generalization of the quantities
$I_1$: the product of a subset of $(2n+1)$ relative distances
squared (see for instance (\ref{I17b})), and $I_2$: the sum of the same subset of relative distances squared (see for instance (\ref{I27b})), are also constants of motion along the trajectory, having a vanishing Poisson bracket with the Hamiltonian. Out of total of $n(2n+1)$ relative distances, there are $n$ of such subsets. They have the meaning of interactions between nearest neighbors, next-to-nearest neighbors and so on:
 {\small
 \begin{align}
 I_1^{(2n+1)} &=
  \begin{cases}
 r_{12}^2 r_{23}^2 r_{34}^2 \ldots   r_{2n,2n+1}^2 r_{1,(2n+1)}^2   \  \, (\mbox{for}\,k_1)\ ,
 \\[3pt]
 r_{13}^2 r_{35}^2 r_{57}^2 \ldots   r_{2n-2,2n}^2 r_{1,2n}^2   \  \, (\mbox{for}\,k_2)\ ,
  \\[3pt]
  \ldots
    \\[3pt]
 r_{1,n+1}^2 r_{n+1,2n+1}^2   \ldots   r_{2n-2,2n}^2 r_{1,2n}^2   \    \, (\mbox{for}\,k_n)\ ,
\end{cases}
\non \\[10pt]
I_2^{(2n+1)} &=
  \begin{cases}
r_{12}^2 +  r_{23}^2 + r_{34}^2 \ldots  +  r_{2n,2n+1}^2 + r_{1,(2n+1)}^2 \   \,   (\mbox{for}\,k_1)\ ,
   \\[3pt]
    r_{13}^2 + r_{35}^2 + r_{57}^2 \ldots   +r_{2n-2,2n}^2 + r_{1,2n}^2   \  \, (\mbox{for}\,k_2)\ ,
    \\[3pt]
      \ldots
    \\[3pt]
 r_{1,n+1}^2 + r_{n+1,2n+1}^2  +  \ldots +  r_{2n-2,2n}^2 + r_{1,2n}^2   \    \, (\mbox{for}\,k_n)\ .
  \end{cases}
\non
\end{align}
}

We conjecture that the product of velocities squared
\begin{equation}
 \tilde{\cal T}\ =\ {\mathbf v}_1^2 {\mathbf v}_2^2 \ldots
   {\mathbf v}_{2n+1}^2 \ =\ \mbox{const\ ,}
   \end{equation}
is also a constant of motion along the trajectory. Note that the quantities $T, \tilde{\cal T}$  are dual to $I_2$ and $I_1$, respectively.

There is also a set of $(2n+1)$ mixed coordinates-velocities quantities
\begin{equation}
  J_i(k) = {\mathbf v}_i^2 + \frac{(k_{1,\ldots n}^2 -1/2)}{(2n+1)^2}  \sum_{\ell < m} b_{\ell,m}^{(i)} r_{\ell m}^2  = 1/2\ ,
  \end{equation}
${\footnotesize i=1,2,\ldots (2n+1)}$  which, we conjecture,  are constants of motion along the lemniscate and are constrained by the relation
\begin{equation}
\sum_{i=1}^{(2n+1)} J_i(k_0) = 2\ {T} + \frac{1}{2n+1}
 (k_{1\ldots n}^2 - \frac{1}{2})   I_{\rm HR}^{(2n+1)}\ .
 \end{equation}

It can be checked that for any $(2n+1)$-body choreography with a certain $k_m(n)$ the total potential is made from two constants of motion, $I_1^{(2n+1)}$ and $
I_{\rm HR}^{(2n+1)}$, thus, it is a superposition the pairwise potentials, has the form
\begin{equation}
\label{Vm}
  {\cal V}_m= \alpha_m  \log {I_1^{(2n+1)}} - \beta_m  I_{\rm HR}^{(2n+1)}\ ,
  m=1, \ldots n\ ,
\end{equation}
cf.(\ref{potential}), (\ref{fivebpotential}), (\ref{sevenpotential}), (\ref{potential7b}), (\ref{potential7c}),
with $\alpha_m=\frac{1}{4}$, independently on $k_m$, while $\beta_m$ takes a certain value which is determined by the requirement that the corresponding choreography is a solution of $(4n)$ coupled Newton equations.

The analysis of the first solution of  (\ref{CMdncondition}), for $z_0$ with
\hbox{$m=1$}, which gives the minimal value of $k$ ($k_{\rm min}=k_1$) and for which the choreography appears with nearest-neighbor interactions (see $I_1^{(2n+1)}$ above), shows that
\[
    \lim_{n \to \infty} k_1^2(n)\  =\ \frac{1}{2}\ .
\]
As $n\to\infty$ the inverted harmonic oscillator potential in (\ref{Vm}) dies out, $\beta_1 \to 0$. In fact, by an explicit calculation of $\beta_1=\beta(k_{min}^2(n))$ in (\ref{Vm}) for the particular cases $n=3,5,7,\ldots 10$ its behavior is very smooth and can be interpolated as
\begin{equation}
\label{betak1asympt}
  \beta(k_{min}^2)  \simeq 0.25091656 (k_{min}^2 -1/2)^{1.4889156}\ .
\end{equation}
It suggests that the leading asymptotic behavior is
\[
\beta(k(n)_{min}^2) \underset{n \to \infty}\ =\ \frac{1}{4}\, \left(k(n)_{min}^2 - \frac{1}{2}\right)^{3/2}\ + \ \ldots \ .
\]
It can be also demonstrated that the limiting value of the period (\ref{period})
for this solution is
\[
 \lim_{n \to \infty} \tau(k_{min}^2(n))\ =\   \frac{\Gamma\left(\frac{1}{4}\right)^2}{\sqrt{\pi}} \  ,
\]
which is twice of the minimal (real) period  of the $\wp$-Weierstrass function with invariants $g_2=1, g_3=0$ (lemniscatic elliptic function). It is also related to the total length of the lemniscate $$l=\frac{1}{\sqrt{2 \pi}}\Gamma \left({\frac{1}{4}}\right)^2\ .$$
In the limit $n \to \infty$ the velocities of all bodies approach the constant value $|{\mathbf v}| = v_{max} = 1/\sqrt{2}$ (see relation (\ref{xvrelation})) as well as the relative distances $r_{i,i+1}$ of the nearest neighbors. The motion of the bodies becomes uniform and the total kinetic energy grows as $T \propto (2n+1)$. Also in the limit $n \to \infty$ the potential (\ref{Vm}) corresponding to minimal $k_1^2\to 1/{2}$ becomes
\[
  {\cal V}_1 \underset{n \to \infty}\ {=}\ \frac{1}{4} \log \left( \prod_{i=1}^{\infty} r_{i,i+1}^2 \right) \ ,
\]
as the results of the fact that the repulsive part of the potential vanishes, \hbox{$\beta_n^{(1)} \rar 0$} (see (\ref{betak1asympt})).
The resulting system becomes a one-dimensional dense Newtonian gas with nearest-neighbor interactions, or, better to say, a one-dimensional {\it Newtonian liquid} moving with constant velocity $v=1/\sqrt{2}$ on a Figure-8 curve - the algebraic lemniscate by Bernoulli. It is remarkable fact that all other interactions - non-nearest neighbors -
die out in this limit. Also, at the large $n$ limit the nearest-neighbor distances become constant: they are equal to length of lemniscate divided by the number of bodies,
\[
  r_{i,i+1} = \frac{l}{2n+1} \ =\ \frac{1}{\sqrt{2 \pi}(2n+1)}\,\Gamma \left({\frac{1}{4}}\right)^2 \ .
\]
As the result, asymptotically, $r_{i,i+1} \propto 1/n$, the density grows $\propto n$ and the potential energy decreases as ${\cal V}_1 \propto -(2n) \ln n$.
Thus, the total energy decays as $E \propto -n \ln n$ (see Fig. \ref{fig11}).

Let us consider the opposite extreme case $k=k^{(max)}=k_n(n)$, when in the limit
\[
     \lim_{n \to \infty}\, k_n(n)^2\  =\ 1 \ ,
\]
but inverted harmonic oscillator potential in (\ref{Vm}) continues to disappear
\[
\beta_n^{(n)} \to 0\ .
\]
As a matter of fact, $k_n(n)$ approaches very fast to the limit $k_n(n)\rar 1$. For instance, for $n=5$, the distance to the asymptotic limit is $\sim 10^{-8}$, while for $n=10$ the distance drops to $\sim 10^{-16}$.

Eventually, the potential (\ref{Vm}) becomes
\[
  V_n \underset{n \to \infty}{=} \frac{1}{4} \log \left( \prod_{i=1}^{\infty} r_{i,i+n}^2 \right) \ ,
\]
which is also a one-dimensional Newtonian gas, with interactions between $n$-distant neighbors $i$ and $i+n$ {\it only}, on the algebraic lemniscate by Bernoulli.
In this case the velocities of bodies tend to zero, the period
\[
 \lim_{n \to \infty} \tau(k_n) = \infty\ ,
\]
and the configuration becomes static. The kinetic energy ${\cal T}$ vanishes as well as $\tilde{\cal T}$.

\begin{figure}
 \includegraphics[width = 2.5in]{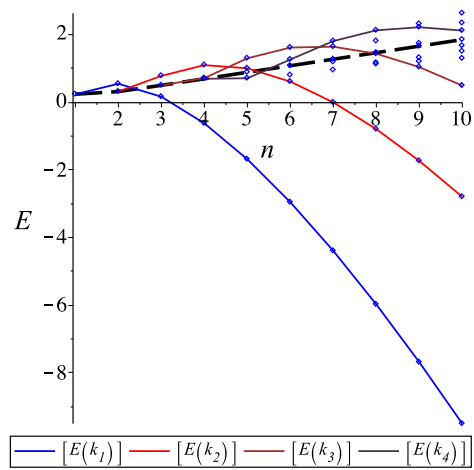}
\caption{\label{fig14}  Energies (marked by dots) $E_{k_m(n)}\ (m=1\ldots n)$ {\it vs} $n$
for the $(2n+1)$-body choreography  for $n=1,2,\ldots , 10$. Lines connecting energies corresponding to $k_1, k_2, k_3$ and $k_4$ drawn.
The dashed-bold line connects the energies corresponding to the maximal value of the elliptic modulus $k=k_n(n)$ for each $n$. }
\end{figure}

Fig.14 summarizes the results of the analysis done for $(2n+1)$-body
choreographies on the algebraic Lemniscate at $n=1,2\ldots 10$.
It shows the possible energies of the system plotted {\it vs} $n$. For each $n$
there are $n$ values of the elliptic modulus $k_1(n) < k_2(n) < \ldots < k_n(n)$
corresponding to $(2n+1)$-body choreographies with fixed center of
mass (see eqs. (\ref{CMdncondition}) and (\ref{nz0})).
For each of those values of $k$ it was checked numerically that the motion is a
solution of the system of $(4n)$ Newton equations corresponding to a potential which is
a combination of two types of pairwise interactions, namely, a repulsive
harmonic oscillator potential between each pair of bodies 
$V_{\rm rep} = - \beta \sum r_{ij}^2$ ($j> i=1,2\ldots 2n+1$), and a logarithmic interaction $V_{\rm log} = \alpha \sum \log r_{ij}^2$ between (i) nearest neighbors interactions only ($j=i+1$ for $k_1$), (ii) between next-to-nearest-neighbors only ($j=i+2$ for $k_2$),
(iii) between next-to-next-nearest-neighbors only ($j=i+3$ for $k_3$) and so on.
It is found that the coefficient $\beta$ in front of the repulsive harmonic oscillator interaction is  decreasing with $n$ and eventually vanishes $\beta_{k_m(n)} \rar 0$ ($m=1\ldots n$) as $n\rar \infty$ (see eq. (\ref{betak1asympt}) for the case $\beta_{k_1(n)}$), while the coefficient $\alpha$ in front of the logarithmic interaction
is found to have the value $\alpha_{k_m(n)} = 1/4$ for all cases ($m=1\ldots n$).  The corresponding energies $E_{k_m (n)} $ {\it vs} $n$  are marked with circles in Fig. \ref{fig14}.  Energies $E(k_{1}(n))$ corresponding to the interaction 
$V_{\rm log}$ among nearest-neighbors  only ($j=i+1$) are joined by a line drawing a parabolic-like curve.  Similarly, and for the sake of exemplification, the energies $E(k_{2}(n))$,  $E(k_{3}(n))$, and $E(k_{4}(n))$ are connected by lines, which correspond, respectively,  to the interaction $V_{\rm log}$ among neighbors $i,j=i+2$ only, $i,j=i+3$ only, and $i,j=i+4$  only. Each of these curves displays a maximum and for large values of $n$ (beyond the maximum) the energies appear ordered $E_{k_1}(n) < E_{k_2}(n) < \ldots $ and going to minus infinity as  $n\rar\infty$ (while the harmonic interaction vanishes in this limit).
Figure \ref{fig14} also shows  that the energies corresponding to  $k_{\rm max}(n)$  grow linearly with $n$.

If T-conjecture is correct, then we expect for all solutions one can find $(8n-1)$ {\it independent} constants of motion (particular Liouville integrals) along the algebraic Lemniscate by Bernoulli. The general scheme hints that the remaining of such constants of motion other than ones indicated above should be in the form of polynomials in the relative distances and velocities with non-integer coefficients. It is beyond the scope of the present work.

\bigskip

\noindent
{\it Conclusions.}
In the present paper, we have shown that each of two 5-body choreographies
with pairwise potentials (\ref{fivebpotential}-\ref{V25b}) (see \cite{JC:2019})
is characterized by 15 explicitly written Liouville integrals which become constants
of motion on the algebraic Lemniscate. Hence, the choreographies are
maximally, particularly superintegrable. The three 7-body choreographies
on the lemniscate were proven to be true choreographies
corresponding to pairwise potentials (\ref{potential7b}), being similar to
the potentials for 3 and 5-bodies cases. We also found a set of Liouville
integrals which have their corresponding counterparts in the cases of 3 and
5-body choreographies.  In particular, the constants
corresponding to the sum and product of certain subsets of  7 relative distances
squared and their dual counterparts, the kinetic energy and the product of the
7 velocities squared, which appears to be a hidden symmetry of the trajectories.
We have analyzed choreographies with up to 21 bodies on the algebraic lemniscate. All of them are found to be solutions of the Newton equations for well defined potentials. Such potentials display a structure similar to that found for the cases of
3,5 and 7 bodies. Namely, at given $n$ for $k_1<k_2<\ldots$ the corresponding potentials contain pairwise logarithmic interaction terms $\log r_{ij}^2$ between neighbors  $i,j=i+1,i=i+2, \ldots$ only together with a repulsive harmonic oscillator interaction between all pair of bodies $(-r_{ij}^2)$. As the number of bodies grows, the repulsive harmonic interaction tends to vanish and in the limit $n \rar \infty$ the system converts to a one-dimensional string with pure logarithmic interaction among equally distant neighbors {\it without collisions}. In particular, for the smallest value $k=k_1$ (which realizes the minimal energy of the system in general),
the interaction occurs between nearest neighbors only, and $k_1^2(n) \rar 1/2$ as $n\rar\infty$, and in this limit the bodies are distributed evenly along the lemniscate, moving  with a constant  velocity (see text above).
We conjecture,  that a $(2n+1)$-body choreography on the lemniscate
with zero angular momentum exists for $n$ different pairwise potentials, defined by $n$ solutions for fixed center of mass and that  all of them are
maximally particularly superintegrable:  it might be an intrinsic property
of the choreographies explaining their existence.
The same phenomenon of the existence of a choreography
manifests the appearance of a new type of equilibrium
configurations: moving, non-steady equilibrium.

It is known that 5-,7-,9-,19-body choreographies on Remarkable
Figure-8-shape trajectory by Moore in ${\bf R}^3$ Newton gravity also exist
\cite{Simo}: the question about their (super)-integrability remains
open.

\bigskip

\noindent
{\it Acknowledgements.} The authors thank T.~Fujiwara, R.~Moeckel and C.~Sim\'o for useful mail correspondence and for personal discussions (T.F. and R.M.).
A.V.T. is grateful to participants of the seminars at mathematics and physics departments, University of Minnesota, Simons Center for Geometry and Physics, C.N. Yang Institute for Theoretical Physics and physics department, Stony Brook University, all at Stony Brook, especially, to V.~Korepin and R.~Schrock for interest to this work.  This research is
partially supported by CONACyT A1-S-17364 and DGAPA IN113819 grants (Mexico).


\end{document}